\documentclass[aip,jmp,author-year,reprint]{revtex4-1}
\usepackage{graphicx}
\usepackage{amssymb}
\usepackage{lineno}
\usepackage{epstopdf}
\usepackage{floatrow}
\usepackage{booktabs}
\usepackage{appendix}
\usepackage{endnotes}
\usepackage{tabularx}
\usepackage{multirow}
\usepackage{hyperref}
\usepackage{dcolumn}
\usepackage{amsmath,amsfonts}
\usepackage{bm}
\usepackage{epstopdf}
\usepackage{placeins}
\newcommand\MyHead[2]{%
  \multicolumn{1}{l}{\parbox{#1}{\centering #2}}
}
\begin{document}
\title{Measurements of high-frequency acoustic scattering from glacially-eroded rock outcrops}
\thanks{Copyright (2016) Acoustical Society of America. A version of this article has been published in J. Acoust. Soc. Am. 139, 1833 (2016) and may be found at \url{http://link.aip.org/link/?JAS/139/1833}. This article may be downloaded for personal use only. Any other use requires prior permission of the author and the Acoustical Society of America.}
\author{Derek R. Olson}
\affiliation{Applied Research Laboratory, Pennsylvania State University, State College, PA, 16804}
\author{Anthony P. Lyons}
\affiliation{University of New Hampshire,
Durham, NH 03824}
\author{Torstein O. S{\ae}b{\o}}
\affiliation{Norwegian Defence Research Establishment,
Kjeller N-2027, Norway}
\date{\today}
%
%
%
%
\begin{abstract}
Measurements of acoustic backscattering from glacially-eroded rock outcrops were made off the coast of Sandefjord, Norway using a high-frequency synthetic aperture sonar (SAS) system. A method by which scattering strength can be estimated from data collected by a SAS system is detailed, as well as a method to estimate an effective calibration parameter for the system. Scattering strength measurements from very smooth areas of the rock outcrops agree with predictions from both the small-slope approximation and perturbation theory, and range between -33 and -26 dB at 20$^\circ$ grazing angle. Scattering strength measurements from very rough areas of the rock outcrops agree with the sine-squared shape of the empirical Lambertian model and fall between -30 and -20 dB at 20$^\circ$ grazing angle. Both perturbation theory and the small-slope approximation are expected to be inaccurate for the very rough area, and overestimate scattering strength by 8 dB or more for all measurements of very rough surfaces. Supporting characterization of the environment was performed in the form of geoacoustic and roughness parameter estimates.
\end{abstract}
\maketitle
\section{INTRODUCTION}

Determination of the relationship between the acoustic field scattered by the seafloor and environmental parameters is crucial to understanding and predicting acoustic interaction with the ocean environment. An important step in this process is to perform measurements of seafloor scattering in conjunction with measurements of seafloor geoacoustic and roughness properties. The scattered field is typically characterized by the differential scattering cross section per unit area per unit solid angle, $\sigma$, which will be abbreviated here as `scattering cross section' or `cross section,' keeping in mind that it is dimensionless. It is a system-independent quantity that characterizes the angular and frequency dependence of the second moment of the acoustic pressure field due to scattering \citep{jackson_richardson_2007,pierce_acoustics}.

In terms of acoustic scattering measurements, rock seafloors have received little attention to date, with five existing scattering strength measurements reported in the literature. To the authors' knowledge, detailed acoustic scattering measurements of rock seafloors, coupled with measured ground truth have never been made. Scattering strength measurements of rock seafloors without quantitative geophysical parameters have been made by \cite{eyering_etal_1948} at 24 kHz, \cite{urick_1954} at 55 kHz, \cite{mckinney_anderson_1964} at 100 kHz, and \cite{soukup_gragg_2003} at 2-3.5 kHz. A report from the Applied Physics Laboratory, University of Washington \citep{APL_9407} presents model curves that were fit to scattering strength measurements taken at a rocky site.

Previous measurements of scattering strength fall between -15 to -22 dB at 20$^\circ$ grazing angle, with the exception of the APL-UW measurement from `rough rock', which is approximately -8 dB at the same angle. These measurements tend to decrease monotonically with grazing angle, apart from typical statistical fluctuations, and some systematic ripples in \cite{soukup_gragg_2003}. Some of these measurements likely suffer from bias. The measurements by \cite{eyering_etal_1948} likely include the effect of multiple interactions with the sea surface and floor, and may represent an overestimate of scattering strength. One of the of the authors of the APL-UW report \citep{jackson_pc_TR9407doubts} has expressed concerns regarding the reliability of the calibration for the measurements on which the models were based. The model curves from the APL-UW report exceed the maximum possible Lambert's law curve, and would violate energy conservation unless the scattering cross section is azimuthally anisotropic or if enhanced backscattering \citep{ishimaru_chen_1990,thorsos_jackson_1991} were present.

The present work addresses the paucity of scattering measurements from rock seafloors by presenting estimates of scattering strength obtained from glacially-eroded rock outcrops, accompanied by characterization of geoacoustic and roughness properties. These outcrops contain two contrasting roughness characteristics that allow model-data comparisons to be made under different conditions. Acoustic backscattering data at 100 kHz were collected off the coast of Sandefjord, Norway by the Norwegian Defence Research Establishment (FFI) aboard the HU Sverdrup II using the HISAS 1030 synthetic aperture sonar (SAS) system from a HUGIN autonomous vehicle \citep{midtgaard_etal_2011,fossum_etal_2008}. This sonar has not been calibrated in terms of its receiver sensitivity $s_r$ or source strength $s_0$, which are required for scattering strength estimates. The product of these two parameters was estimated by comparison of measured data to a model which used measured input parameters. Roughness estimates of rock outcrops were obtained using a digital stereo photogrammetry system. Geoacoustic parameters were estimated using an effective medium model with previously measured mineral composition of the bedrock in the area.

Characterization of the environment, including estimates of geoacoustic and roughness parameters of the bedrock is summarized in Section~\ref{sec:envchar}. Section~\ref{sec:expOver} gives an overview of the acoustic scattering experiment, and details the data processing and calibration technique. Scattering strength results are presented and discussed in Section~\ref{sec:results}, with conclusions given in Section~\ref{sec:conclusions}.

\section{ENVIRONMENTAL CHARACTERIZATION}
\label{sec:envchar}
\subsection{Geoacoustic characterization}
\label{sec:geoacoustic}
The bedrock surrounding Larvik and Sandefjord, Norway and their coastline is composed of monzonite, a crystalline intrusive igneous rock \citep{petersen_1978, neumann_1980, lemaitre_2002}. This material supports both compressional and shear waves, and like most natural materials, contains intrinsic dispersion and attenuation \citep{mavko_rockPhysics}. Wave propagation within monzonite has been modeled as an elastic medium with frequency-independent complex wave speeds. This model contains transverse and longitudinal waves, but is dispersionless with linear dependence of attenuation on frequency. Since linear attenuation over a limited frequency range results in logarithmic dispersion via the Kramers-Kronig relations it does not satisfy causality \citep{futterman_1962,milton_etal_1997}. Although the model is acausal, it is commonly used to characterize the seafloor especially when little information is available \citep[Chap. 9]{jackson_richardson_2007}. The frequency-dependence of interface scattering is primarily dependent on the roughness spectrum, and is weakly dependent on sound speed dispersion. Thus including appropriate dispersion in the elastic model would likely not be observable when comparing scattering measurements to models.

Parameters for the elastic model are $\tilde{c}_{p}$ and $\tilde{c}_t$, the compressional and shear phase speeds, $\delta_p$ and $\delta_t$, the compressional and shear loss parameters (imaginary component of the complex sound speed divided by the phase speed), and $\rho_b$, the bulk density. The complex wave speeds, $c_p$ and $c_t$, are related to the phase speeds by $c_p = \tilde{c}_p (1+i \delta_p)$ and $c_t = \tilde{c}_t (1+i \delta_t)$ respectively, where $i$ is the imaginary unit \cite[Chap. 9]{jackson_richardson_2007}. The bulk and shear moduli, $K$ and $\mu$ respectively, are related to $\tilde{c}_p$ and $\tilde{c}_t$ through the standard formulae: $\tilde{c}_p = \sqrt{(K+ \frac{4}{3} \mu)/\rho_b}$ and $\tilde{c}_t = \sqrt{\mu/\rho_b}$ \citep{mavko_rockPhysics}.

Geoacoustic parameters of the area were not measured, but bounds were computed by using an effective medium approximation combined with previously measured mineral compositions \citep{neumann_1976, neumann_1980}. Crystalline igneous rock is composed of randomly-oriented crystal grains of individual minerals \citep{lemaitre_2002}. An effective medium model is used to replace the heterogeneous granular material with a homogeneous material with properties that reflect the aggregate effect of the crystal grains. The narrowest bounds for the aggregate bulk and shear moduli without knowledge of grain shapes or distributions are attained by the Hashin-Shtrikman-Walpole (HSW) bounds for multiphase composites \citep{mavko_rockPhysics}. For each mineral component, its bulk modulus, shear modulus, density, and volume fraction $\beta_i$ are required. The two isotropic moduli are computed from anisotropic crystalline mineral elastic properties using a Voigt average \citep{denToonder_etal_1999} and the bulk density is computed using a simple volume average.

Due to the slight porosity of crystalline igneous rock, one of the components of the effective medium is water, the volume fraction of which was not measured by \cite{neumann_1976, neumann_1980}. Measurements of porosity in granite, a similar crystalline igneous rock, were made by \cite{tullborg_2006} in Sweden and by \cite{norton_knapp_1977} in the United States. These measurements range between 0.61\% and 2.60\% with a mean of 1.02\% and a standard deviation of 0.43\%. The porosity mean is used to compute the mean wave speeds and the bulk density, and the standard deviation is used to compute model uncertainties.
 
Attenuation at 100 kHz cannot be estimated using the HSW bounds. Instead, an estimate for attenuation is based on measurements of water-saturated granite measured by \cite{coyner_martin_1990}. Using the resonant bar technique, the measured quality factor at 100 kHz, the center frequency of the HISAS 1030, was approximately 30. This estimate results in $\delta_p \approx 0.02$. Shear wave attenuation in saturated granite was found by the same researchers to be approximately twice that of compressional wave attenuation in this frequency range, so $\delta_t = 2 \delta_p$. 

The isotropic averaged elastic moduli ($K$ and $\mu$), density ($\rho$), and volume fraction ($\beta$) for each mineral in the Sandefjord area can be found in Table \ref{tab:mineralParameters}, with the results of the Voigt average for isotropic bulk and shear moduli, and density. Inputs to the Voigt average for each mineral are from \cite{lbRFM}. These moduli were used to estimate HSW bounds on compressional velocity, shear velocity and density, and are reported in Table \ref{tab:geoacousticParameters}, along with the attenuation parameters. Since detailed information regarding mineral composition is rarely available for a given region, Table~\ref{tab:geoacousticParameters} also includes tabulated parameters for generic rock from Table IV.2 in \cite{APL_9407}, and generic granite (a similar crystalline igneous rock) from Table 5.2 in \cite{bourbie_1987_acoustics}.  The granite parameters are the mean of ranges reported in \cite{bourbie_1987_acoustics}, which are of 4500 - 6000 m/s for $c_p$, 2500 - 3300 for $c_t$, and 2500 - 2700 kg/m$^3$ for $\rho_b$. The parameters for generic rock generally perform quite poorly, although parameters for generic granite are much closer to the estimates using detailed minerology. This comparison indicates that tabulated values for geoacoustic properties may be adequate for modeling purposes if the lithology is known.

\begin{table}
\centering
\begin{tabular}{ l r r r r}
\hline \hline
Mineral Name & $K$ [GPa] & $\mu$ [GPa] & $\rho$ [kg /m$^3$] &$\beta$ [\%] \\ 
\hline 
Water & 2.20 & 0.00 & 1000 & 1.20 \\ 
Albite & 79.17 & 22.14 & 2630 & 51.05 \\ 
Anorthite & 110.84 & 42.88 & 2760 & 8.97 \\ 
Orthoclase & 81.71 & 31.85 & 2560 & 23.91 \\ 
Nepheline & 80.43 & 32.86 & 2620 & 3.32 \\ 
Diopside & 166.31 & 65.77 & 3310 & 3.45 \\ 
Enstatite & 184.70 & 75.99 & 3200 & 0.08 \\ 
Olivine & 129.60 & 81.05 & 3224 & 2.88 \\ 
Magnetite & 161.00 & 91.30 & 5180 & 2.05 \\ 
Apatite & 125.80 & 49.65 & 3218 & 1.35 \\ 
\hline \hline
\end{tabular}
\caption{Mineral composition of the experimental site, along with their volume fractions, $\beta$, isotropic bulk modulus, $K$, isotropic shear modulus, $\mu$, and density, $\rho$. Mineral fractions have been averaged from three sites in \cite{neumann_1976,neumann_1980}, and elastic constants used to estimate isotropic moduli are from \cite{lbRFM}.}
\label{tab:mineralParameters}
\end{table} 
\begin{table}
\centering
\begin{tabular}{ l c c c c c c}
\hline \hline
\noalign{\smallskip}
Parameter &  \MyHead{1.3cm}{Lower\\ Bound} & \MyHead{1.3cm}{Upper\\ Bound} & Mean & \MyHead{1.5cm}{Generic\\Rock }& \MyHead{1.5cm}{Generic\\Granite} \\
\noalign{\smallskip}
\hline
$\tilde{c}_p$ [m/s] & 5945 & 6842 & 6393 & 3600 & 5300\\
$\tilde{c}_t$ [m/s] & 3198 & 3353 & 3276 &  1900 & 2900\\
$\rho_b$ [kg/m$^3$] &  2696 & 2720 & 2708& 2500 & 2600\\
$\delta_p$ & 0.02 &0.02 & 0.02 & 0.0018 & 0.01\\
$\delta_t$ & 0.04 & 0.04 & 0.04 & 0.085 & 0.05\\
\hline \hline
\end{tabular}
\caption{Summary of geoacoustic parameter estimates and their bounds. The values for $\tilde{c}_p$, $\tilde{c}_t$ and $\rho_b$ were calculated using the Hashin-Shtrikman-Walpole bounds \citep{mavko_rockPhysics}. Also included are estimates for generic rock from Table IV.2 in \cite{APL_9407}, and average values for generic granite (which is similar to monzonite) from Table 5.2 in \cite{bourbie_1987_acoustics}}
\label{tab:geoacousticParameters}
\end{table}
\subsection{Roughness characterization}
\label{sec:roughness}
An experiment to characterize the roughness of rock outcrops was performed in May 2013 near Sandefjord, Norway at 59$^\circ$4'26.2''N, 10$^\circ$15'42.1''E. Roughness measurements of in-air (subaerial) glacially-eroded rock outcroppings called \textit{roches moutone\'es} were obtained using digital stereo photogrammetry. These roughness measurements provided inputs to the effective acoustic system calibration described below, and inputs to approximate scattering models that were compared with measured data. These types of outcrops have two contrasting roughness characteristics: a gently-undulating surface where the ice flowed onto the outcrop (stoss), and a stepped surface where the ice flowed off of the outcrop (lee). The stoss side has been shaped through the mechanism of glacial clast abrasion, whereby sediment grains (clasts) trapped beneath glaciers gouge and scrape the underlying bedrock \citep{scholz_1976,alley_1997}. The resulting surface is characterized by large-scale undulations that follow the glacial flow pattern, and small-scale scratches, or striae, from individual clasts or sediment grains. The stepped leeward side is formed when hydraulic fracturing dislodges blocks of rock delineated by the internal joint structure, a process termed glacial quarrying or plucking \citep{hallet_1996,iverson_2012, zoet_anandakrishnan_alley_2012}.

\begin{figure}
\centering
\includegraphics[width=\columnwidth]{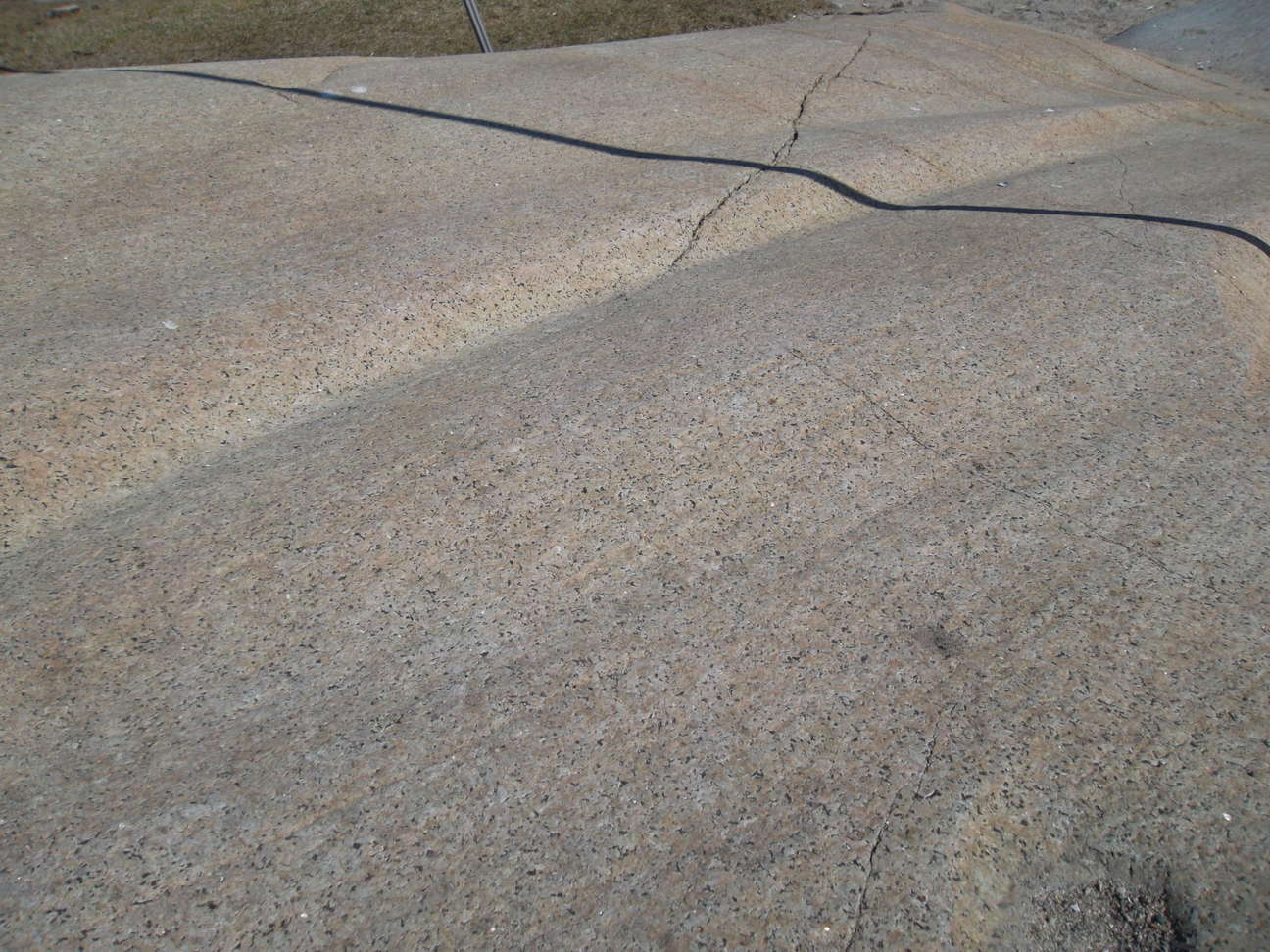}
\caption{(color online) Photograph of in-air (subaerial) glacially abraded roughness measured in air. Ice flow in this area was from the bottom left of the image to the top right. The large-scale  undulations can be seen, and are represented in the height fields in Fig.~\ref{fig:abradedHeights}. The distance between the surface shown in the bottom left and upper right is approximately 2.5 m. The diagonal line in the upper part of the image is a shadow cast by the measurement system.}
\label{fig:area4photo}
\end{figure}

\begin{figure}
\centering
\includegraphics[width=\columnwidth]{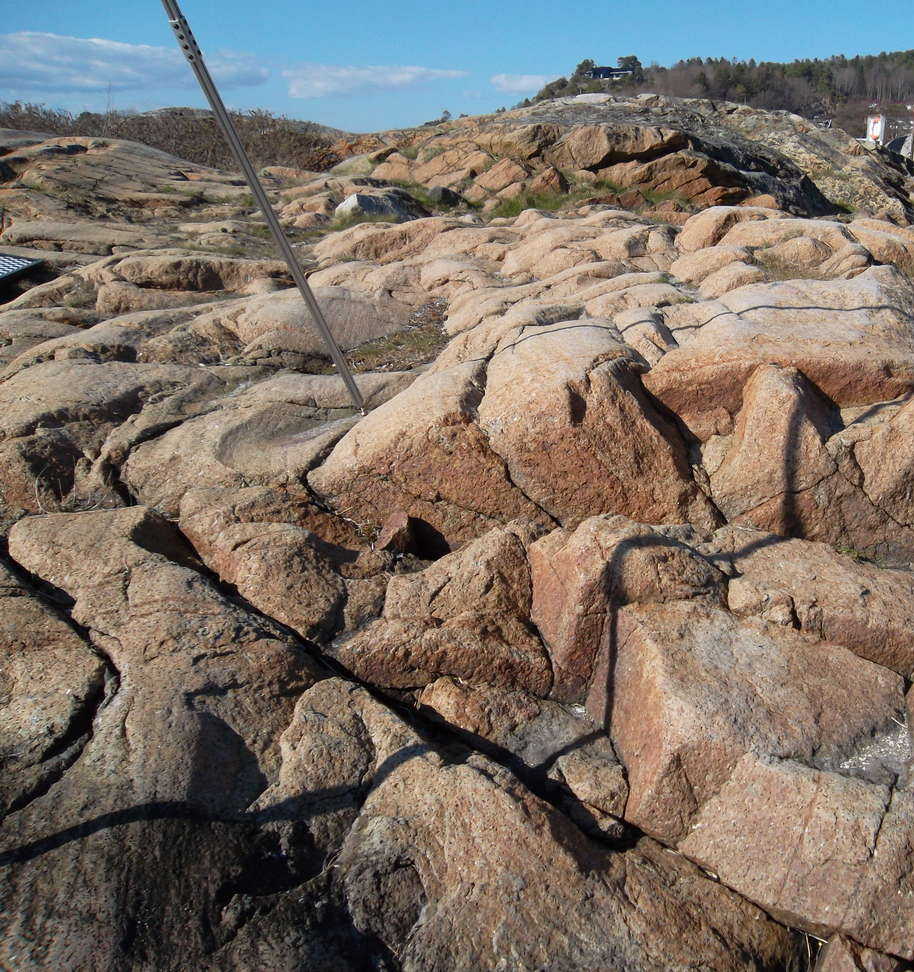}
\caption{(color online) Photograph of glacially plucked roughness. Ice flow in this area is out of the page. The typical stepped characteristics of the leeward side of a \textit{roche moutone\'e} can be seen here. One leg of the stereophotogrammetry frame can be seen in the upper left quadrant of the image, of which 1.3 m is visible.}
\label{fig:area5photo}
\end{figure}

Roughness measurements are reported from two in-air surfaces that contain glacial abrasion and plucking respectively. They were chosen because they are representative of other surfaces with the same geomorphology in the region. To illustrate the roughness characteristics of these surfaces, photographs are shown in Fig.~\ref{fig:area4photo} for glacial abrasion, and Fig.~\ref{fig:area5photo} for glacial plucking. These surfaces are the in-air expression of the same geomorphological features studied below in Section~\ref{sec:expOver}. Similarities with the SAS imagery of these features is explored below.  The ice flowed from North to South at the roughness measurement  site \citep{mangerud_2004}. Additional measurements and analysis can be found in \cite{olson_thesis}.

In this research, acoustic measurements were performed on submerged outcrops, which preserved the glacial features. Roughness measurements were performed on subaerial outcrops, which are subject to additional erosion through chemical weathering \citep{nicholson_2009} over the approximately 12,000 years since the glaciers retreated from this area \citep{mangerud_2011}. Chemical erosion creates small pits in the rock surface as grains are dissolved due to oxidation and exposure to an acidic environment. These features are the primary source of difference between roughness characteristics of submerged and subaerial \textit{roches moutone\'es}. It is argued below that the chemical weathering pits do not affect the scattered acoustic field at these frequencies, and that subaerial roughness measurements are acceptable as input parameters to scattering models used to predict the cross section from submerged outcrops.

The stereo photogrammetry system used in these measurements consisted of two Nikon D7000 digital single lens reflex cameras with Nikkor 28 mm f/2.8 D lenses mounted to an aluminum frame. Baseline separations were set at approximately 0.5 m and 1 m, which enabled the system to operate at heights of both 1 m and 2 m from rock surfaces, which will be called the high- and low-resolution modes respectively. The nominal system resolutions determined by the camera resolution, camera separation, focal length, and mean distance from the rock surface are $(\Delta x, \Delta y, \Delta z) = ( 160, 158,77.4)$  $\mu$m for the high-resolution mode and $( 356, 357,176)$ $\mu$m for the low-resolution mode, where $\Delta x$ is the resolution parallel to the baseline direction in the horizontal plane, $\Delta y$ is the resolution perpendicular to baseline in the horizontal plane, and $\Delta z$ is the precision of the depth estimate. Since a 7$\times$7 correlation window was used for the stereo-matching algorithm in the photogrammetry processing, a conservative estimate of the realistic image-plane resolutions is seven times the nominal resolution, (1.12,1.11) mm and (2.49,2.45) mm for the high- and low-resolution modes respectively. Surface features as small as the nominal system resolution are observable, but features at scales smaller than the realistic resolution are subject to an effective low-pass filter due to the correlation window. These resolutions correspond to Nyquist-Shannon spatial frequencies of approximately 449 m$^{-1}$ and 202 m$^{-1}$ for the high- and low-resolution modes respectively. The cameras were calibrated using a black and white checkerboard pattern attached to a glass plate. Images were processed to obtain height fields using the OpenCV library \citep{openCV}, and the intrinsic and extrinsic camera parameters were obtained using the camera calibration toolbox \citep{bouguet_camCalBox}. Details on stereo photogrammetry can be found in \cite{manualPhoto} and an example of an underwater system can be found in \cite{lyons_pouliquen_2004}.

\subsubsection{\textbf{\textit{Data analysis}}}
Two-dimensional roughness power spectra were estimated from measured height fields using Thomson's multitaper approach \citep{thomson_1982,percival_walden}. This approach uses the discrete prolate spheroidal sequences (DPSS) as orthogonal window functions. One of the advantages of the DPSS windows is that for a given value of the equivalent noise bandwidth, $N_{BW}$, several orthogonal windows are available. Power spectra computed with orthogonal windows were incoherently averaged to produce a power spectrum estimate with reduced variance compared to single realizations \citep{percival_walden}. Two-dimensional window functions were obtained by taking the inner product of two one-dimensional window functions. To further mitigate spectral leakage, a least-squares plane was subtracted from the height field before windowing and spectrum estimation. The $NBW$ parameter was set at six, and seven windows in each direction were used, for a total of 49 orthogonal windows.
\subsubsection{\textit{Roughness results}}
\begin{figure}
\centering
\includegraphics[width=\columnwidth]{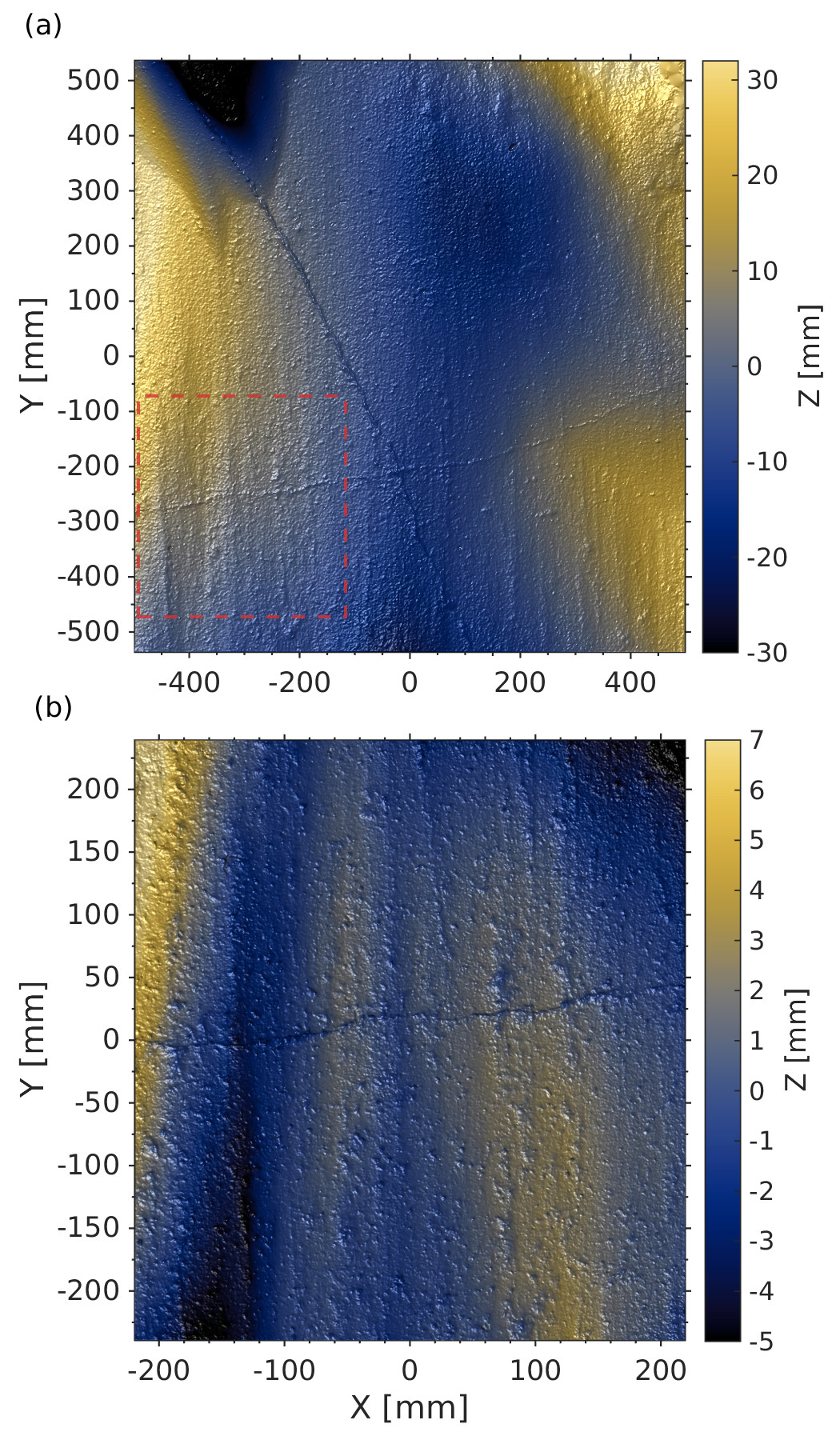}
\caption{(color online) Rough interface results from a glacially abraded surface in (a) the low-resolution mode, and (b) the high-resolution mode. The glaciers flowed in the negative $y$ direction. The color bar corresponds to height reference to the surface mean, and the brightness, or black/white information communicates the surface slope. The dashed box in (a) indicates the portion of the surface that is shown in (b).}
\label{fig:abradedHeights}
\end{figure}

Two roughness measurements of a glacially abraded rough surfaces are presented in Fig.~\ref{fig:abradedHeights}. These measurements were made of the same rock surface, but at the two different heights of the roughness measurement system, 2 m for Fig.~\ref{fig:abradedHeights}(a), and 1 m for Fig.~\ref{fig:abradedHeights}(b). A mean plane was subtracted from each of the measurements before plotting. Because of this operation it is difficult to use a global coordinate system for all roughness measurements, and coordinates are referenced to their mean values. The region of overlap between the two measurements is indicated by the dashed box in Fig.~\ref{fig:abradedHeights}(a).

Roughness is displayed in a color representation in which the color bar communicates surface height, and surface slope information modifies the black/white balance (lightness), as if the surface were illuminated by a light source. The lightness modification to the color bar is biased to lighter values, since black is part of the color bar and white is not. Thus pixels may appear to have colors that are not part of the color bar (e.g. white glints) due to a large modification of its lightness value. This visualization scheme is used because the dynamic range of the color scale is dominated by the large amplitude features and cannot resolve the low-amplitude, short wavelength features captured by the photogrammetry system.

After a mean plane was subtracted, the root mean square (rms) height of Fig.~\ref{fig:abradedHeights}a is 13.1 mm and the rms slope is 0.32. In Fig.~\ref{fig:abradedHeights}b, the rms height is 2.01 mm and the rms slope is 0.37.  In both images, glacier flow is approximately in the negative $y$ direction. Long wavelength undulations in both the along- and across-flow direction are evident at scales of approximately 50 cm. At wavelengths on the order of a few cm, the roughness is primarily caused by scratches parallel to the glacier flow direction. These striae are likely caused by individual clasts being dragged across the rock surface by the glacier. At wavelengths less than 5 mm there exist pits in the surface that are likely the result of post-glacial chemical weathering \citep{nicholson_2009} that do not seem to be shaped in any preferred direction.

Post-glacial weathering is the largest source of discrepancy between submerged and subaerial bedrock roughness characteristics. From Fig.~\ref{fig:abradedHeights}(b), weathering pits are common, although their prevalence is exaggerated somewhat by the shading scheme. There are also several small cracks running through both height fields. The cracks are likely pre-existing joints in the rock material that were widened by freeze-thaw cycles \citep{nicholson_2009}. Although pits 10 mm wide and 1 mm deep exist, they are relatively rare, with most pits less than 2-3 mm in horizontal extent, and shallower than 0.5 mm. It is likely that roughness parameters estimated for wavelengths larger than 2-3 mm are applicable to underwater (submereged) abraded surfaces. Since first-order perturbation theory \citep{thorsos_jackson_1989} states that roughness at scales less than $\lambda$/2 (7.5 mm for the HISAS 1030 sonar) cannot affect the scattered field, and perturbation theory is expected to be accurate for these surfaces, chemical weathering pits likely do not affect the scattered field.

\begin{figure}
\centering
\includegraphics[width=\columnwidth]{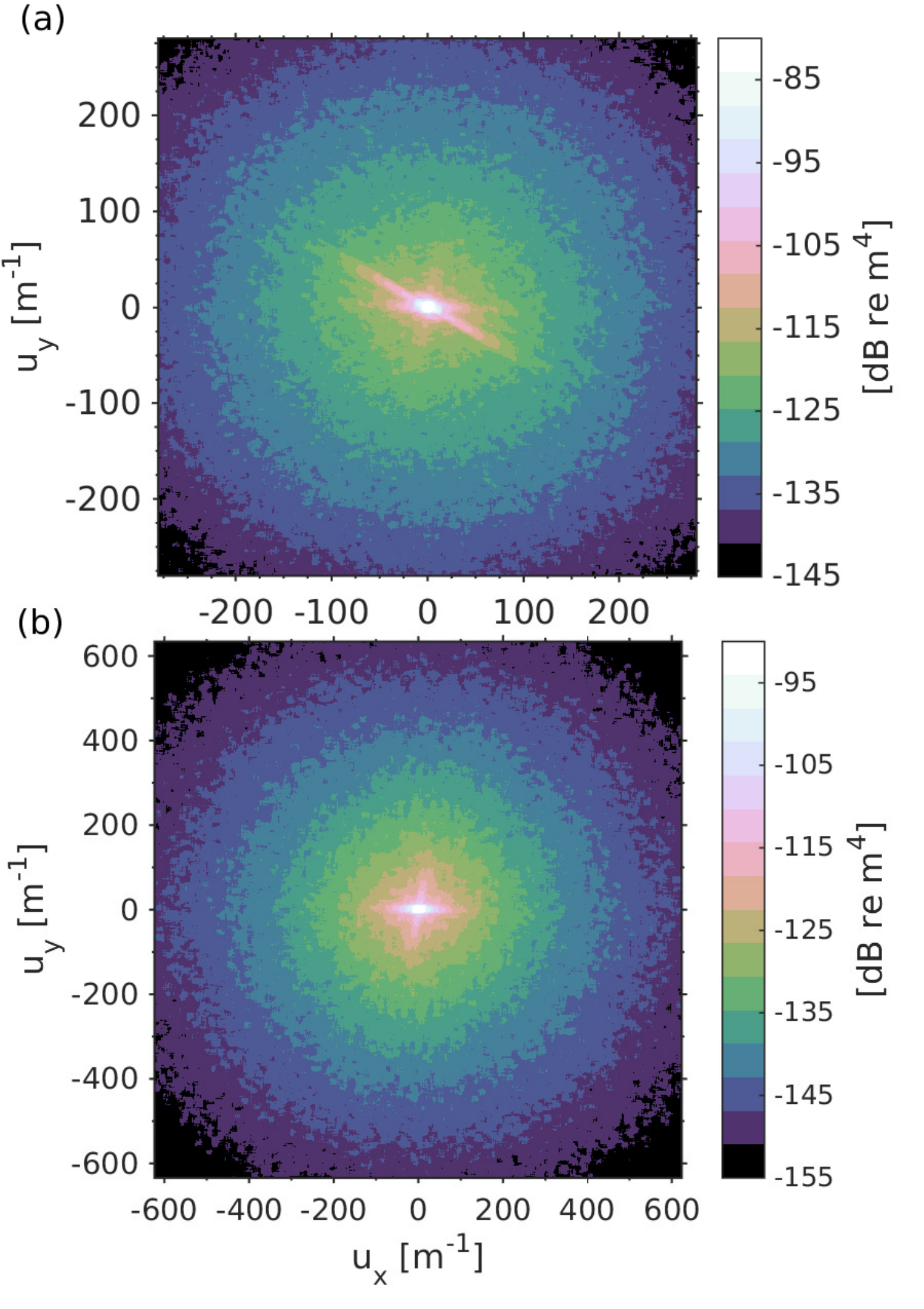}
\caption{(color online) Decibel version of the two dimensional power spectra in m$^4$ of abraded surfaces shown in Fig.~\ref{fig:abradedHeights} as a function of horizontal spatial frequencies, $u_x$ and $u_y$. The power spectrum of the abraded surface measured in low-resolution mode is presented in (a), and the spectrum measured by the high-resolution mode is in (b). Angles mentioned in the text are measured counterclockwise from the +$u_x$ axis.}
\label{fig:abradedSpectra}
\end{figure}

Power spectra estimated for the height fields depicted in Fig. \ref{fig:abradedHeights} are presented in Fig.~\ref{fig:abradedSpectra}. The 2D power spectra are plotted as a function of horizontal spatial frequencies, $u_x$ and $u_y$. Both spectra are anisotropic at low spatial frequencies, but are isotropic at high spatial frequencies with the division at approximately 200 m$^{-1}$. This division between large and small scales corresponds to a 5 mm wavelength. Since the chemical weathering pits appear to be isotropic and are mostly restricted to diameters of less than 5 mm, the high-spatial frequency isotropic regime would seem to represent the effect of chemical weathering. Since the weathering pits are confined to a region of spatial frequencies above the highest Bragg spatial frequency accessible by the HISAS sonar (135 m$^{-1}$), they likely do not contribute significantly to the scattered field. So long as attention is restricted to spatial frequencies less than 200 m$^{-1}$, the spectral characteristics of subaerial roughness can be used to represent spectral characteristics of submerged \textit{roches moutone\'es}.

The low-wavenumber anisotropy takes the form of broad peaks in the spatial frequency domain centered at the origin, and at aspect angles, $u_\phi=\tan^{-1}(u_y/u_x)$, of 0$^\circ$, -21$^\circ$, and 77$^\circ$ for Fig~\ref{fig:abradedSpectra}(a), and 0$^\circ$ and 77$^\circ$ for the Fig.~\ref{fig:abradedSpectra}(b). By the projection-slice theorem \citep{ferguson_wyber_2005}, angles in the spatial-frequency domain correspond to angles in the spatial domain. Since these peaks are centered at the origin, their widths set the correlation scale of the surface in that particular direction. The width of the anisotropic peak at 0$^\circ$ likely corresponds to the correlation scale of small scratches perpendicular to the direction of glacier flow. The anisotropic feature at 77$^\circ$ is present in both measurements and likely corresponds to large-scale undulations approximately parallel to glacier flow. The peak at -21$^\circ$ in Fig.~\ref{fig:abradedSpectra}(a) likely corresponds to the undulations present between the upper left and lower right corners of Fig.~\ref{fig:abradedHeights}(a).

\begin{figure}
\centering
\includegraphics[width=\columnwidth]{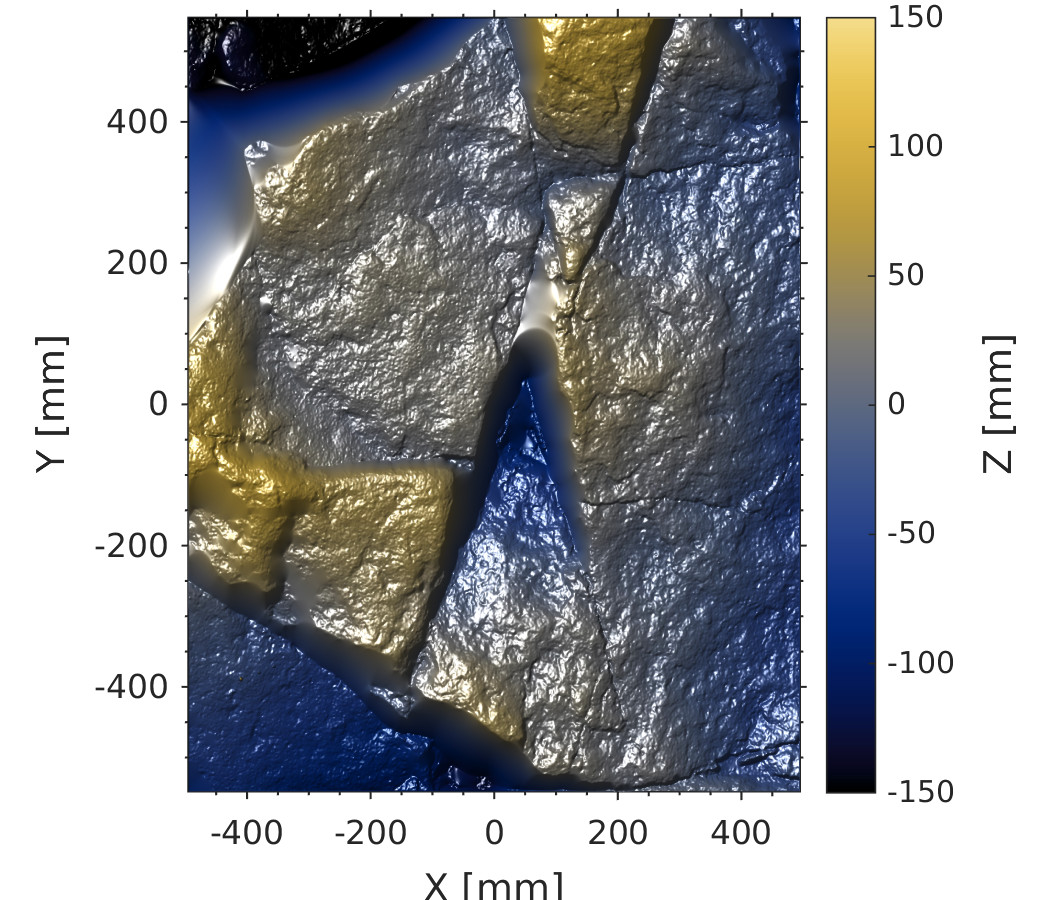}
\caption{(color online) Roughness measurement results for the plucked rough interface. The colorscale communicates surface height, and the surface slope has been included as changes to grayscale value to accentuate low-amplitude roughness not resolved by the color scale. Only the low resolution measurement is presented because the high resolution mode does not include enough facets to obtain a proper sample size.}
\label{fig:pluckedSurf}
\end{figure}

A rough surface formed through glacial plucking is presented in Fig.~\ref{fig:pluckedSurf} with the same visualization scheme as in Fig.~\ref{fig:abradedHeights}. This surface is composed of approximately planar polygonal facets at large scales, with low-level roughness superimposed at small scales. After a mean plane was subtracted, the rms height of this surface was 45.6 mm, and the rms slope is 5.2. Note that some of the steep faces appear to be quite smooth. This artifact results from a shortcoming of wide-baseline photogrammetry in which the stereo correspondence algorithm fails for steep slopes, and the missing areas are interpolated. The small-scale roughness appears to be isotropic, and lacks the parallel striae exhibited by glacially abraded surfaces. It is likely that the small-scale roughness reflects the shape of the preexisting internal joint surface before glacial quarrying, and is not the result of glacial abrasion \citep{iverson_2012}.

\begin{figure}
\centering
\includegraphics[width=3.375in]{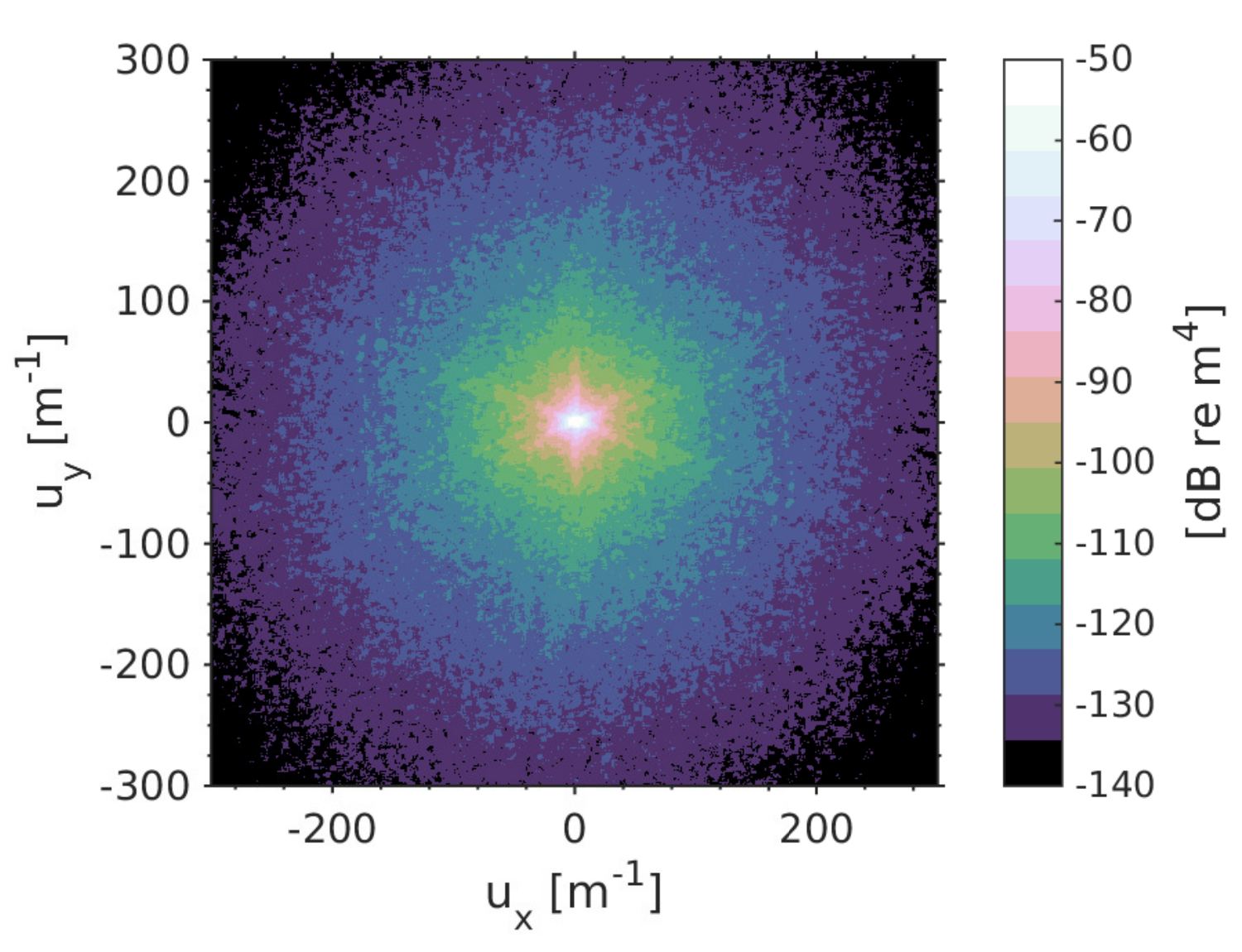}
\caption{(color online) Decibel version of the two-dimensional power spectrum in m$^4$ of the plucked interface shown in Fig.~\ref{fig:pluckedSurf} as a function of horizontal spatial frequencies, $u_x$ and $u_y$. Angles mentioned in the text are measured counterclockwise from the +$u_x$ axis.}
\label{fig:pluckedSpectrum}
\end{figure}

The two-dimensional power spectrum of this surface is shown in Fig.~\ref{fig:pluckedSpectrum}. The plucked spectrum is anisotropic over much of the spatial frequency domain. At low spatial frequencies, it has peaks at $u_\phi$ of 90$^\circ$, and $\pm$23$^\circ$. These directional peaks have most of their energy at the origin and extend to spatial frequencies of 150 m$^{-1}$. These peaks likely result from the large-scale facet structure of the plucked interface. Aside from the directional peaks, there is a background isotropic spectrum that decays as a power law, likely representing the isotropic small-scale roughness on each facet.

\begin{figure}
\centering
\includegraphics[width=\columnwidth]{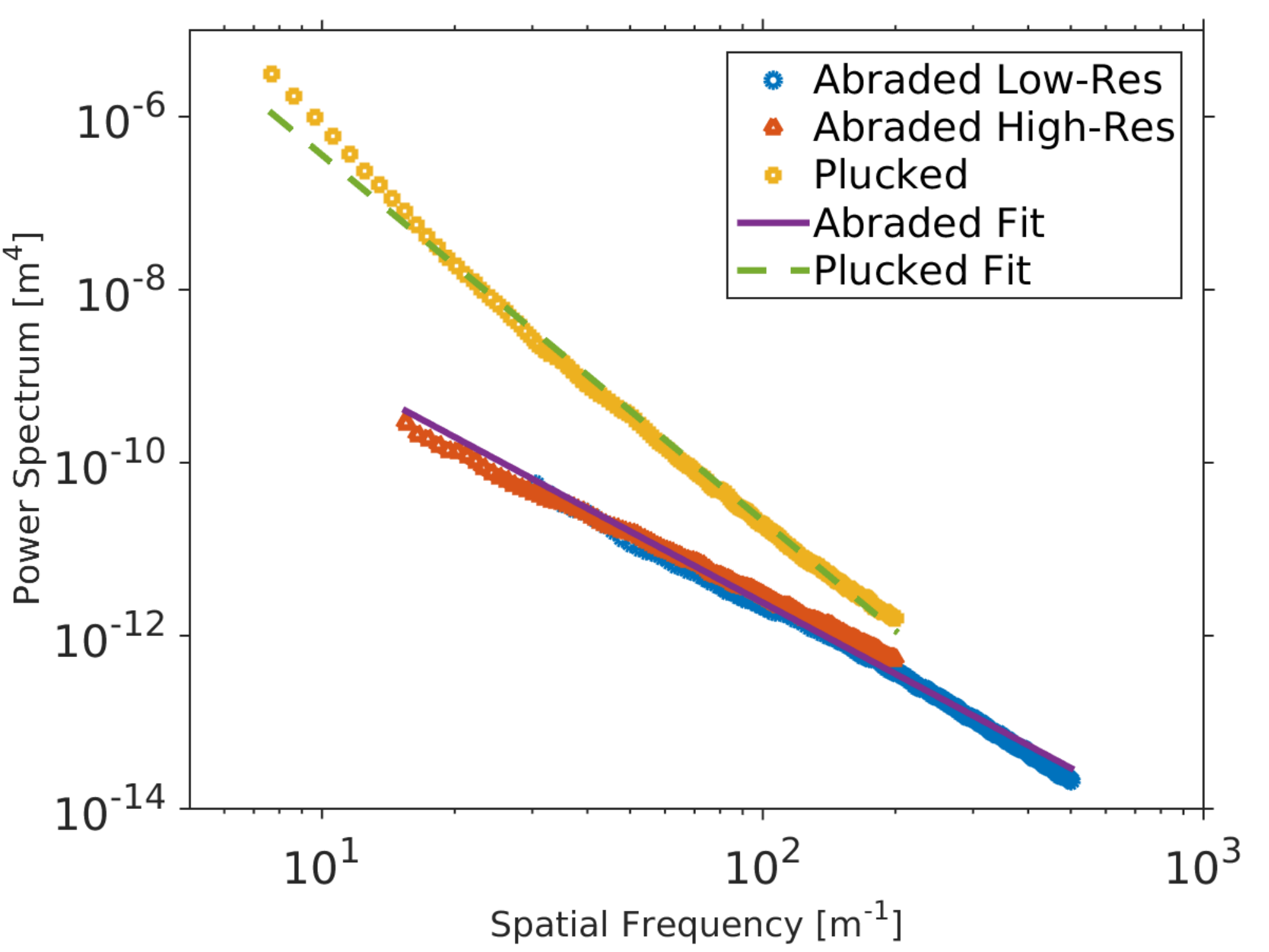}
\caption{(color online) Azimuthally averaged power spectrum from abraded and plucked surfaces with power-law fit. The power spectrum from both the low- and high-resolution modes of the photogrammetry system are plotted for the abraded surfaces.}
\label{fig:abradedSpectrumAzAvg}
\end{figure}

Parameters for an isotropic two-dimensional power spectrum are required for the effective acoustic system calibration step detailed below, and as inputs for scattering models. Averaging over azimuth is typically only performed for isotropic spectra. In this case the azimuthally averaged spectrum is expected to reflect the behavior of scattering cross section measurements averaged over several azimuth angles, which may be anisotropic. Based on perturbation theory, the roughness spectrum components responsible for backscattered power are at the Bragg wavenumbers, $2 k_w \cos\theta$, or spatial frequencies or $2 u_w \cos\theta$, where $k_w= 2\pi u_w = 2\pi f/c_w$ is the wavenumber in water, $c_w$ is the sound speed in water, and $f$ is the acoustic frequency. For the angles covered by the experimental geometry, this corresponds to spatial frequencies between 105 m$^{-1}$ and 135 m$^{-1}$. The azimuthally averaged spectrum for the small- and large-scale abraded spectrum, and the plucked spectrum are shown on a log-log scale in Fig.~\ref{fig:abradedSpectrumAzAvg}. Low wavenumbers that are biased by the apodization functions are not shown here. The small- and large-scale abraded spectra match very closely in their overlapping spatial-frequency domains, which is expected. The plucked spectrum exceeds the abraded spectra by more than three orders of magnitude in the low spatial frequency domain, but is much closer in power at high spatial frequencies.

A power-law model is fit to azimuthally-averaged spectra, of the form
\begin{align}
\Phi(u_r) = \phi_2 / u_r^{\gamma_2}
\end{align}
where $\Phi(u_r)$ is the 2D power spectrum, $u_r$ is the radial spatial frequency, $\phi_2$ is the spectral strength, and $\gamma_2$ is the spectral slope. The subscripts on $\gamma_2$ and $\phi_2$ indicate that these parameters are for 2D power spectra, following the convention in \cite{jackson_richardson_2007}. Power-law fits are displayed in Fig.~\ref{fig:abradedSpectrumAzAvg}, and estimated parameters can be found in Table~\ref{tab:roughnessParams}. All spatial frequency components of the spectrum that were not biased from apodization or the photogrammetry processing were used in the fit. Although a single power-law is fit to the plucked spectrum, it appears to have some curvature in log-log space and could converge with the abraded spectra at high spatial frequencies if higher-resolution data were available.

\begin{table}
\centering
\begin{tabular}{ l l l}
\hline \hline
Parameter & Abraded & Plucked\\
\hline
$\phi_2$ [m$^{4-\gamma_2}$] & 4.847$\times10^{-8}$ & 6.083$\times10^{-3}$\\
$\gamma_2$ & 2.73 & 4.36\\
$\eta_\Phi$ [$\%$] & 4.5 & 6.2\\
\hline \hline
\end{tabular}
\caption{Roughness Spectrum model parameters and uncertainty.}
\label{tab:roughnessParams}
\end{table}

Uncertainty in the model parameter estimates is a major contributor to uncertainty in the resulting scattering strength measurements reported below. Parameters are estimated using a least-squares fit to the power-law in log-log space, which means that parameter uncertainty is a function of the residual sum of squares, and number of points used in the estimate. An additional source of uncertainty is the variation of $\Phi$ as function of azimuthal spatial frequency, $u_\phi$. The total relative variance $\eta^2_\Phi$ in the spectrum estimate is estimated by $\eta^2_\Phi = \eta^2_{LS} + \eta^2_{aniso}$, where $\eta^2_{LS}$ is the relative variance due to the least squares fit, and
\begin{align}
\eta^2_{aniso} = \frac{1}{u_{max} - u_{min}}\int\limits_{u_{min}}^{u_{max}}\frac{\langle \Phi^2 (u_\phi,u_r)-\Phi^2(u_r)\rangle_{u_\phi}}{\Phi^2(u_r)}\,\textrm{d} u_r
\end{align}
represents the azimuthal variability in the power spectrum. It is the variance of the power spectrum over $u_\phi$ divided by the squared mean, and averaged over the Bragg spatial frequencies accessible by the measurement system. A parameter for total uncertainty of the spectrum over the Bragg spatial frequencies is used rather than uncertainty of individual parameter estimates. Uncertainties can be found in Table~\ref{tab:roughnessParams}.
\section{ACOUSTIC SCATTERING EXPERIMENTS}
\label{sec:expOver}
Acoustic backscattering measurements were performed off the coast of Larvik and Sandefjord, Norway in April 2011 \citep{midtgaard_etal_2011}. This experiment was performed by the Norwegian Defence Research Establishment (FFI) aboard the HU Sverdrup II. Data were collected using the HUGIN 1000 HUS autonomous underwater vehicle (AUV) equipped with the HISAS 1030 interferometric SAS system \citep{fossum_etal_2008}. This sonar has not been calibrated in terms of its open-circuit receiver sensitivity, $s_r$, or the source strength, $s_0$, although beam patterns were measured. These parameters must be determined in order to estimate the absolute seafloor scattering cross section.

The seafloor in this area consisted of \textit{roches moutone\'es} surrounded by a sediment of cobble with a mud matrix. The water depth at the experimental site was approximately 30 m, with nominal vehicle altitude of 10 m from the seafloor. The sound speed profile was slightly upward-refracting with the surface sound speed at approximately 1452 m/s, and approximately 1456 m/s at the seafloor. The change in sound speed of the lower 10 m of the water column was a maximum of 3 m/s, and therefore refraction effects on the local grazing angle can be ignored. Further details on the experiment can be found in \cite{midtgaard_etal_2011}.
\subsection{Synthetic aperture sonar overview}
\label{sec:sasOver}
Synthetic aperture sonar (SAS) is a data collection and beamforming technique in which a transmitter and a receiver array move along a track, which is typically linear or circular. Acoustic energy is transmitted at regular intervals, and the resulting scattered field is sampled, also at regular intervals. The received field from many transmissions can be concatenated to form a synthetic array that has a length many times that of the physical receiver array. A SAS image can be formed using a variety of beamforming techniques, the simplest and most robust of which is the backprojection, or delay and sum algorithm \citep{hawkins_thesis}. Synthetic array lengths scale linearly with pixel range and images have a constant Cartesian resolution over the whole scene. The array lengths can be quite long and in most situations imaged objects are in the near field. Aligning successive physical aperture locations to form a synthetic array is challenging in the ocean environment due to hydrodynamic forces on the sonar platform. Typically, an inertial navigation system combined with displaced-phase center acoustic navigation is used to estimate the sonar element locations to within a fraction of a wavelength \citep{bellettini_pinto_2002}. More information on SAS processing can be found in \cite{hansen_sas} and references therein. 

Interferometric SAS is possible with two vertically-separated receiver arrays on a sonar platform. The phase difference between SAS image pixels formed from each array is related to the seafloor height. Since phase is wrapped to $2\pi$ radians, discontinuities in the phase difference must be detected in the presence of noise. Typically, an $n\times n$ window is used to provide an estimate seafloor height with reduced variance at the cost of reduced resolution \citep{saebo_etal_2013, saebo_thesis}. Bathymetry estimates can be used in scattering strength measurements by providing an estimate of the local seafloor slope and global grazing angle at each pixel. The assumption of a constant or planar seafloor is often used in scattering strength measurements \citep{jackson_etal_1986}, which is severely violated in the case of the rock outcrops studied in this research. The local seafloor slope is estimated using a weighted average version of a finite difference operator on the SAS bathymetry to reduce high-frequency noise. The weights are computed using an $m$-point least squares fit to a quadratic function, with the derivative of the polynomial computed analytically \citep{savitsky_golay_1964}. Second-order polynomials were used to estimate all slopes, with window sizes of 55 pixels for plucked surfaces, and 23 pixels for abraded surfaces. The larger window size was used to suppress effects of steps on leeward surfaces of \textit{roches moutone\'es} and focus on the mean trend.

\subsection{Estimating the scattering cross section from synthetic aperture sonar data}
\label{sec:dataProc}
Estimation of scattering strength from SAS systems is similar in principle to estimation using other real-aperture sonar systems designed to measure scattering strength \citep{jackson_etal_1986,williams_etal_2002}, with the exceptions of nearfield beamforming and the ability to estimate seafloor slope. Typically, beamformed time series are incoherently averaged over independent areas of the seafloor, and then the sonar equation is inverted for scattering strength for each intensity sample. Since SAS is a near-field imaging technique, transmission loss changes significantly along the synthetic array for all ranges. Consequently, sonar equation terms that vary as a function of sensor location, i.e. spherical spreading and attenuation, are removed before the image is beamformed. Terms that do not vary as a function of sensor position, such as the ensonified area and calibration parameters, are removed after image formation. The resulting calibrated SAS pixel intensity values correspond to the unaveraged cross section, $\tilde{\sigma}$, for azimuthally isotropic scatterers.

A sonar equation is used that relates pixel intensity to the scattering cross section. It is equivalent to Eq. (G.11) in \cite{jackson_richardson_2007} but adapted to SAS quantities. This equation is valid so long as the scattering cross section and vertical beam pattern are both slowly varying within the system resolution. Let $v_{ij}$ be the complex matched-filtered voltage sensed by the $j$th receiver element, and delayed such that it corresponds to the $i$th pixel. Let $x$ be the along-track position of the pixel location, and $y$ be the cross-track (ground range) position. The complex value of the $i$th pixel, $q_i$, is the output of the delay-and sum beamformer, and is a weighted sum over the synthetic array, with the weights determined by transducer patterns, and by tapering applied to reduce side-lobes. When corrected for propagation, vertical beam pattern effects, and coherent gain, $q_i$ is defined by
\begin{align}
\begin{split}\label{eq:pressureToPixel}
q_i &= \left(\sum\limits_{j}^{Nr} w_j b_{tx}(\phi_{ij}) b_{rx}(\phi_{ij})\right)^{-1}\times \\
 &\left( \sum\limits_{j}^{Nr} w_j b_{tx}(\phi_{ij}) b_{rx}(\phi_{ij}) \frac{r^2_{ij}e^{2 \alpha r_{ij}}}{\left| b_{tx}(\theta_{ij}) b_{rx}(\theta_{ij})\right|}v_{ij} \right)
\end{split}
\end{align}
where $w_j$ is the processing weight applied to the $j$th receiver, $\alpha$ is the attenuation of seawater, $r_{ij}$ is the distance (slant range) from pixel $i$ to receiver $j$, $b_{tx}(\phi_{ij})$ and $b_{rx}(\phi_{ij})$ are the transmit and receive horizontal beam patterns, and $b_{tx}(\theta_{ij})$ and $b_{rx}(\theta_{ij})$ are the vertical beam patterns. The variables $\phi_{ij}$, and $\theta_{ij}$, are the horizontal and depression angles from the sensor $j$ and pixel $i$ in the sonar's coordinate system. The HISAS 1030 sonar has a single transmitter and multiple receivers. The phase-center approximation \citep{bellettini_pinto_2002} is employed to work with an equivalent monostatic configuration with colocated transmitters and receiver. The unaveraged scattering cross section, $\tilde{\sigma}$, is related to $q_{i}$ through
\begin{align}
\left | q_i \right |^2 = (s_r s_0)^2 \Gamma \tilde{\sigma} \delta x \delta y \frac{\cos(\theta - \theta_0)}{\cos \theta}
\label{eq:pixelToSigma}
\end{align}
where $s_r$ is the receiver voltage sensitivity in V/Pa, $s_0$ is the source level in Pa-m, and the cosine terms result from converting from the transducer array coordinate system to the seafloor coordinate system, with $\theta_0$ the depression angle of the main response axis of the transmitter as discussed in Appendix G of \cite{jackson_richardson_2007}. The quantities $\delta x$ and $\delta y$ are the along track and range resolutions of the system respectively. These resolutions are defined as the width of rectangular windows that have the same integrated power as the ambiguity function along a particular direction, similar to the concept of equivalent noise bandwidth. This definition of resolution differs from others, such as the 3 dB down points, and is a simple way to define a sonar equation. The local seafloor slope at each pixel is used to define the local grazing angle. Ensembles are formed by grouping local grazing angles into 1$^\circ$ bins and estimating the sample mean.

The along-track resolution, $\delta x$, is nominally 3.14 cm for the parameters of the HISAS 1030 and processing parameters used for these data. Due to vehicle motion, the value of $\delta x$ fluctuates by as much as 5\% based on calculations of the along track resolution for each pixel. To simplify data processing, the nominal value of $\delta x$ is used, and its variability is included in uncertainty analysis. Note that since SAS is a nearfield imaging algorithm, a given patch of the seafloor is always in the nearfield of the synthetic array. The Fourier transform relationship between the beam pattern in the far field and the aperture weighting function is used to compute $\delta x$, since the beam pattern in the focal plane of a focused near-field array is identical to the far-field beampattern \citep{mast_2007}. The range resolution, $\delta y$ is determined by the transmitted bandwidth and spectral weighting of the received pulse, and is equal to 3.25 cm.

The relative signal gain, $\Gamma$, is defined here as the power ratio between the partially coherent array gain and the coherent array gain \citep{cox_1973,carey_moseley_1991,carey_1998}. It characterizes the bias observed when the received signal is a fluctuating quantity instead of a point scatterer. In this work, partial coherence is due to two mechanisms, 1) phase fluctuations due to uncertainty in the sensor positions and oceanographic conditions, and 2) amplitude and phase fluctuations due the scattering characteristics of random rough surfaces. The first mechanism is included as a source of uncertainty and discussed at the end of the next subsection.

The spatial coherence of rough surface scattering is a complex topic and a rigorous treatment is outside the scope of this work. However, an intuitive argument is given that $\Gamma$ is the same for all pixels. In general, the coherence of the field due to scattering from a rough surface has a Fourier transform relationship with the covariance of the pressure at the scattering surface \citep{mallert_fink_1991}. Theoretical models of coherence typically employ an isotropic point scatterer model (see \cite{jackson_moravan_1984} and references therein), or the van Cittert-Zernike theorem (vCZT) \citep{born_wolf, mallert_fink_1991}. Under the experimental conditions in this work, both models predict that the coherence along the receiver aperture is equal to the autocorrelation of the transmitting aperture function. This relationship, which we will call the vCZT, depends on the two equivalent assumptions that the covariance of the surface field behaves like a Dirac delta function, or that the scattering cross section is isotropic. For the ground ranges studied in this work, the grazing angles along the synthetic array vary by less than 0.1$^\circ$. Therefore we can restrict attention to azimuthal coherence, and only the aziumthal dependence of the scattering cross section is of import. Since azimuthally isotropic scatterers have already been assumed above, it is an appropriate assumption to use for coherence as well. Since the scattering cross section is not anisotropic for all situations, this result may not always be applicable. However, the intuitive argument can be made that if the bistatic scattering cross section is isotropic over the azimuth angles subtended by the transmitter and receiver arrays, then the spatial coherence is practically equal to the autocorrelation of the transmitter aperture.

Using the vCZT, $\Gamma$ can be demonstrated to be the same for all pixels. The transmit and receive arrays are always matched in length due to the phase center approximation \citep{bellettini_pinto_2002}. Additionally, the transmit and receive aperture weights have functional forms that scale as a function of $x/y=\tan(\phi)$, where $\phi$ is the horizontal angle from the sensor location to a given pixel. Therefore range dependence in both the partial coherent gain and coherent gain is canceled out by the definition of $\Gamma$. Under the assumption of azimuthal isotropy, $\Gamma$ is identical for all pixels and may be normalized out during the calibration procedure described below. Note that if the system had been calibrated using a technique such as reciprocity, the value for $\Gamma$ would need to be explicitly calculated based on the system parameters.

\subsection{System calibration}
\label{sec:sysCal}

\begin{figure*}
\centering
\includegraphics[width=0.9\textwidth]{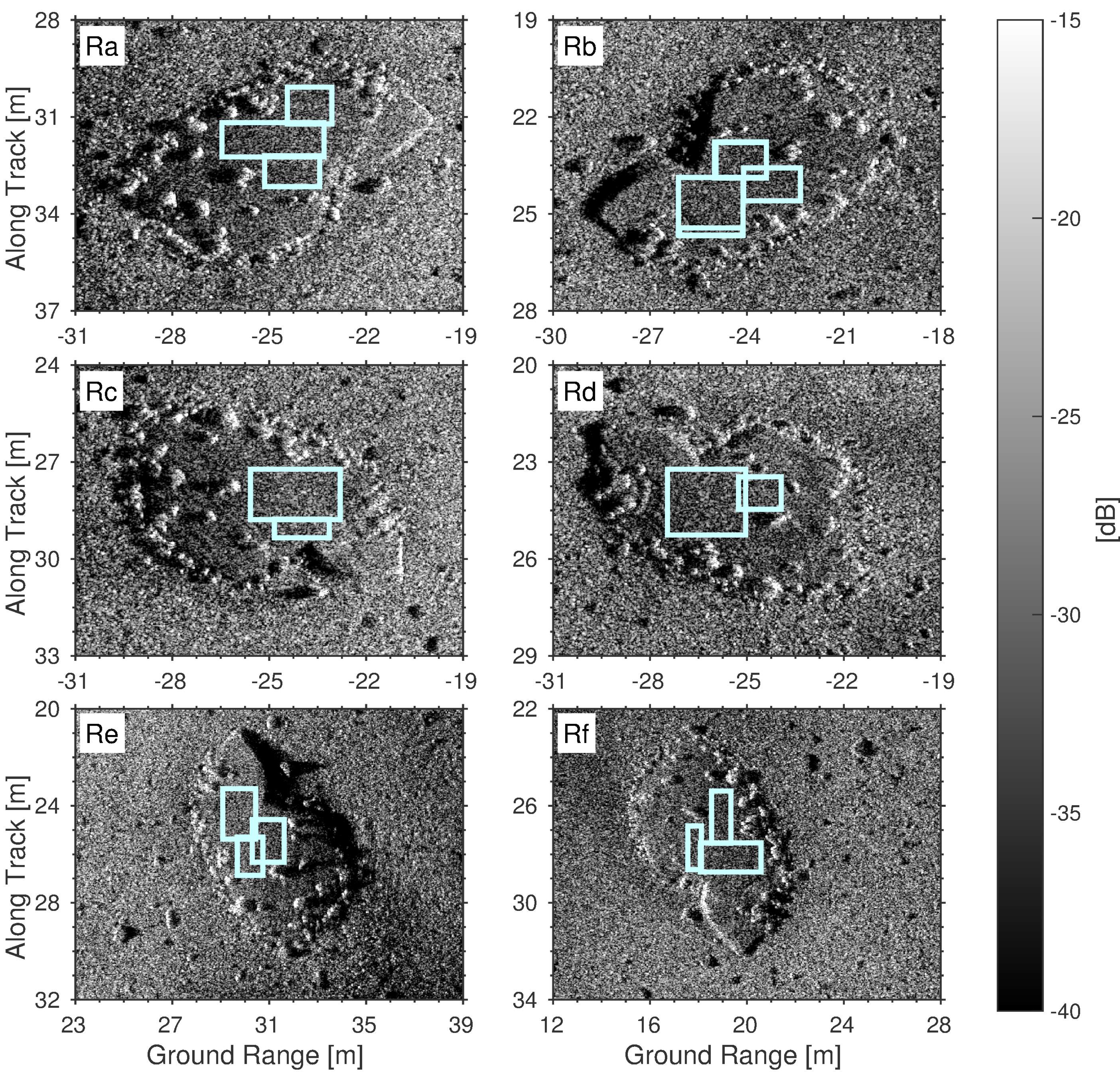}
\caption{(color online) SAS images of the calibration rock outcrop. Boxes denote areas where pixels were extracted to form estimates of scattering strength. With respect to Ra, the rest of the images (Rb-Rf in order) are related in azimuth angle by the following clockwise rotations: 180$^\circ$, 45$^\circ$, 225$^\circ$, 270$^\circ$, and 90$^\circ$. Grayscale value denotes the decibel equivalent of $\tilde{\sigma}$, the unaveraged scattering cross section. Horizontal axes represent ground range from the sonar track with positive values representing the port side, and negative values representing the starboard side. Vertical axes represent distance along the sonar track.} 
\label{fig:SAScalibrationRock}
\end{figure*}

The two parameters characterizing system calibration, $s_r$ and $s_0$, have not been measured, but are required for estimating the scattering cross section. An effective calibration technique is applied that estimates, $s = (s_r s_0)^2$, by fitting the scattering strength of an area of the seafloor to a valid scattering model with known inputs. Note that this technique also normalizes out any bias that is constant for all pixels and measurement locations, such as the relative signal gain, $\Gamma$, discussed above.  This method of calibration has been applied previously using a seafloor scattering model by \cite{dwan_thesis} for the Gloria side-scan sonar, and used for scattering strength measurements by \cite{lyons_anderson_1994}. A similar technique using man-made targets, such as metal spheres, has become standard technique \citep{foote_etal_2005} for sonar calibration.

A rock outcrop with very low-amplitude roughness was chosen based on inspection of SAS imagery to provide an effective system calibration. The apparent low-amplitude roughness enabled the use of a first-order scattering model. SAS images of the calibration rock taken at various azimuthal angles are presented in Fig.~\ref{fig:SAScalibrationRock}. This outcrop extends approximately 0.5 m above the surrounding sediment. The color scale corresponds to the decibel value of the relative scattering cross section. Note that the decibel quantity here and in Fig.~\ref{fig:sasSvabergs} is dimensionless and thus appears without a reference unit. The vertical axes correspond to the distance along the synthetic aperture, and horizontal axes correspond to the ground range to the sonar, with negative values representing distance from the port side of the vehicle, and positive values from the starboard side. Boxes indicate areas that were used to estimate scattering strength. In the case that multiple boxes intersect, their union is used to define a more complicated boundary that avoids double counting of pixels. Ensonification direction is nominally from the line where the ground range is zero, and can be from the left or right of the images depending on the sign of the ground range values. The calibration rock outcrop is approximately 10 m x 5 m in horizontal extent, and appeared to be flat and smooth based on the image texture, as supported by interferometric SAS bathymetry. Dropstones, likely deposited by glaciers, are present on the rock surface. Apart from discrete features such as dropstones and edges, the image amplitude appears to slowly decrease as the absolute value of the ground range from the sonar track increases. Since all propagation and system effects have been removed, this slow change in image amplitude can be attributed to the decrease in scattering strength with grazing angle, which is confirmed in the scattering strength plots below. The surrounding sediment is composed of approximately 4 cm cobble in a mud matrix, as indicated by diver samples. The surface of the calibration rock scatters sound at approximately 7 dB lower than surrounding cobble, which is a further indication that the surface is quite smooth. The calibration rock surface, like most of the exposed bedrock in this area, has been formed through glacial abrasion, as indicated by the smooth image texture and low amplitude of the scattered field.

The relative scattering strength is averaged over all of these aspects for the system calibration. In \ref{sec:roughness}, the roughness power spectra of glacially abraded surfaces was seen to be anisotropic at scales on the order of the Bragg wavenumber for this sonar system. To minimize bias due to anisotropy, the calibration procedure compares azimuthally-averaged scattering strength to the small-slope model computed with azimuthally-averaged roughness power spectrum parameters. Azimuthal variability of the roughness spectrum at the range of Bragg spatial frequencies was used to compute error bars on the scattering model, and thus the estimate of $s$. 

The first-order small-slope approximation (SSA) \citep{voronovich_1985} was used to compute a model cross section curve from which $s$ is estimated. In its first-order form, the SSA can be decomposed into the product of a term depending on the geoacoustic properties, and the Kirchhoff integral \citep[Chap. 13]{ jackson_richardson_2007}. The Kirchhoff integral was numerically evaluated using the algorithm developed by \cite{gragg_etal_2001} and \cite{drumheller_gragg_2001}. The SSA is accurate in the respective regions of validity of the simpler small-roughness perturbation approximation and Kirchhoff approximation \citep[Chap. 13]{jackson_richardson_2007}. Although these bounds have not been established for power-law surfaces, perturbation theory for fluid interfaces and a von K{\'a}rm{\'a}n spectrum has been shown to be accurate with $k_w h\approx 1$, where $h$ is the rms roughness \citep{thorsos_jackson_williams_2000}. The rms roughness of the surface presented in Fig.~\ref{fig:abradedHeights}(b) is 2.09 mm, which corresponds to $kh = 0.86$, and the rms slope is 0.37. Given these considerations, the SSA is expected to be accurate for glacially abraded surfaces.

The elastic SSA model requires 9 inputs, $\phi_2$, $\gamma_2$, $c_p$, $c_t$, $c_w$, $\delta_p$, $\delta_t$, $\rho_b$, and $\rho_w$, the density of water. It was argued in \ref{sec:roughness} that roughness characteristics of the surfaces from which the acoustic calibration measurements were made are similar to the roughness characteristics of glacially abraded surfaces at the roughness measurement site. Therefore, the power spectrum estimates presented in \ref{sec:roughness} are assumed to represent the roughness power spectrum of the calibration rock. Since the two sites are approximately 5 km apart, the roughness characteristics are not exactly the same. However, since many clasts have been dragged across the surface, small-scale roughness is assumed to be well-approximated by a stationary random field. The statistics of this random height field can then be assumed to be the same at both sites, given that the mineral composition and erosion mechanisms are the same. Mean geoacoustic properties and their bounds were summarized in Table~\ref{tab:geoacousticParameters}.

\begin{figure}
\centering
\includegraphics[width=\columnwidth]{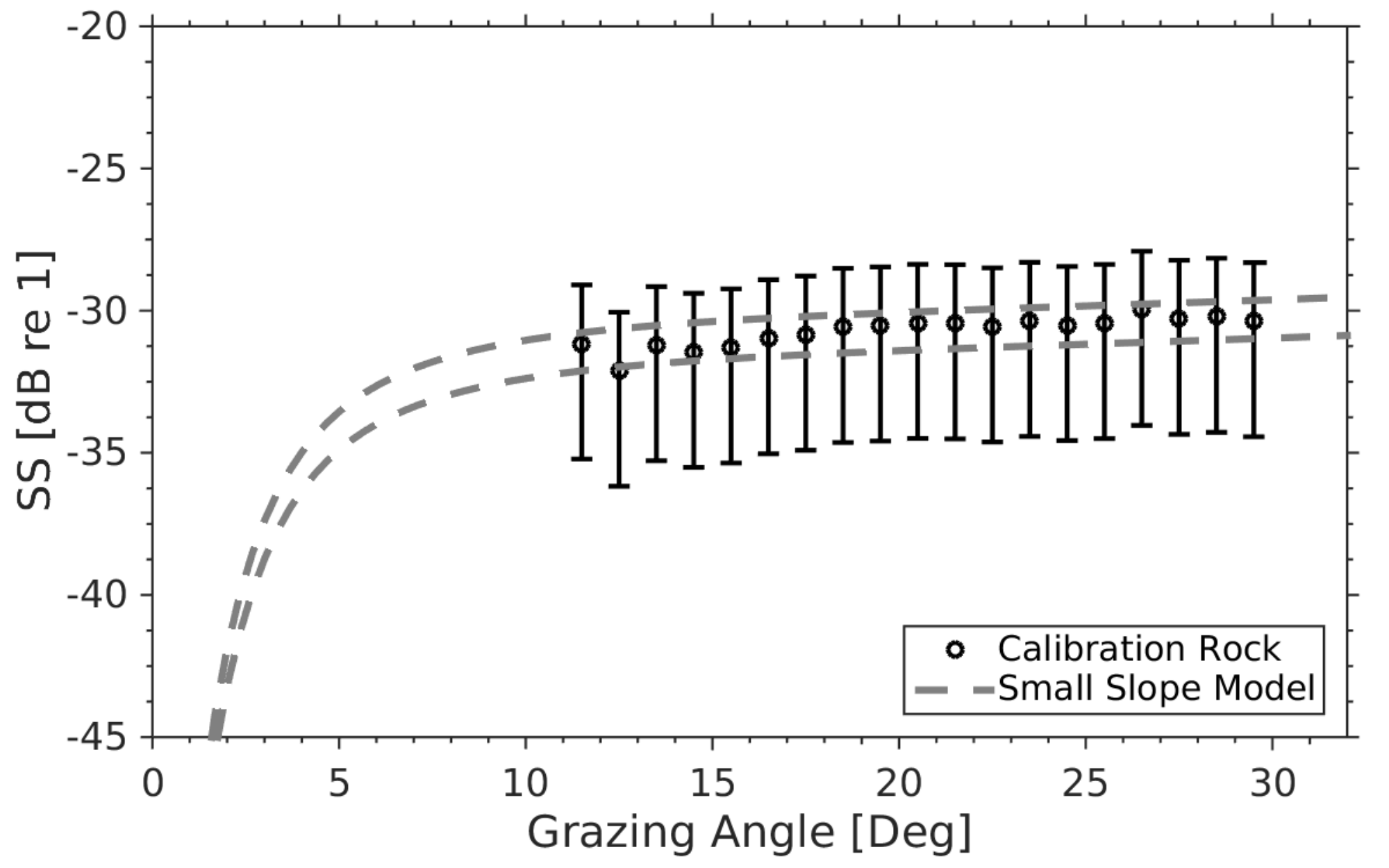}
\caption{Results from the effective calibration technique. Shown here is the scattering strength as a function of grazing angle averaged over all azimuths of the calibration rock and using estimated value of $s$. It is compared to the SSA model with dashed lines representing the model uncertainty. Uncertainties for measured data are represented as error bars.}
\label{fig:calibration}
\end{figure}

The effective calibration parameters, $s$, is estimated through a least-squares fit of model cross section, $\sigma_{SSA}$, to the measured cross section, $\sigma$ over the angular domain covered by the data. A comparison between $\sigma_{SSA}$ and the measured scattering strength from the calibration rock using the best-fit value of $s = 7.03\times10^{-6}$ V$^2$m$^2$ is presented in Fig.~\ref{fig:calibration}. The measured data and SSA model curves are quite similar in shape over the grazing angles of interest. These data exhibit no dramatic minima or maxima, indicating that there is no critical angle in the angular domain of the measurement. From the geoacoustic parameters listed in Table.~\ref{tab:geoacousticParameters}, the compressional and shear wave critical angles are 77$^\circ$ and 63$^\circ$ respectively. Below the critical angle, the shape of the scattering strength curve is insensitive to the compressional and shear wave speeds contrast, and more sensitive to the density contrast.

Uncertainty in scattering strength estimates is a function of uncertainty in $s$, $\delta y$ and the finite ensemble of pixels used to estimate the mean scattered power. Ensemble sizes for the abraded surfaces reported below were between 220 and 737 samples with a mean of 561 samples. Ensemble sizes for the plucked surfaces were between 1148 and 2525 samples with a mean of 2109 samples. Uncertainty in $s$ is driven by the finite ensemble from the calibration rock, and uncertainty in the model parameters used to compute $\sigma_{SSA}$ and $\delta y$. Imperfect knowledge of vehicle motion and the ocean environment causes the array to become partially coherent, as discussed above. At present, there is no technique to estimate the degree of phase error along the synthetic aperture \citep{hansen_saebo_2013}, so it will be assumed that the element positions are known to within $\lambda$/10, a common requirement for well-focused images \citep{hansen_etal_2011}. This requirement translates into a residual coherence bias of $1- \exp\lbrace -\pi^2/50\rbrace$, or 18\%. This bias is included as an additional relative uncertainty contribution because the navigational accuracy is expected to vary from site to site and cannot be normalized out by the calibration technique. The total uncertainty of scattering cross section estimates, expressed as 95\% confidence intervals, is approximately 61\%, or 6 dB (+2 dB, -4 dB).

\section{RESULTS AND DISCUSSION}
\label{sec:results}

\begin{figure*}
\centering
\includegraphics[width=\textwidth]{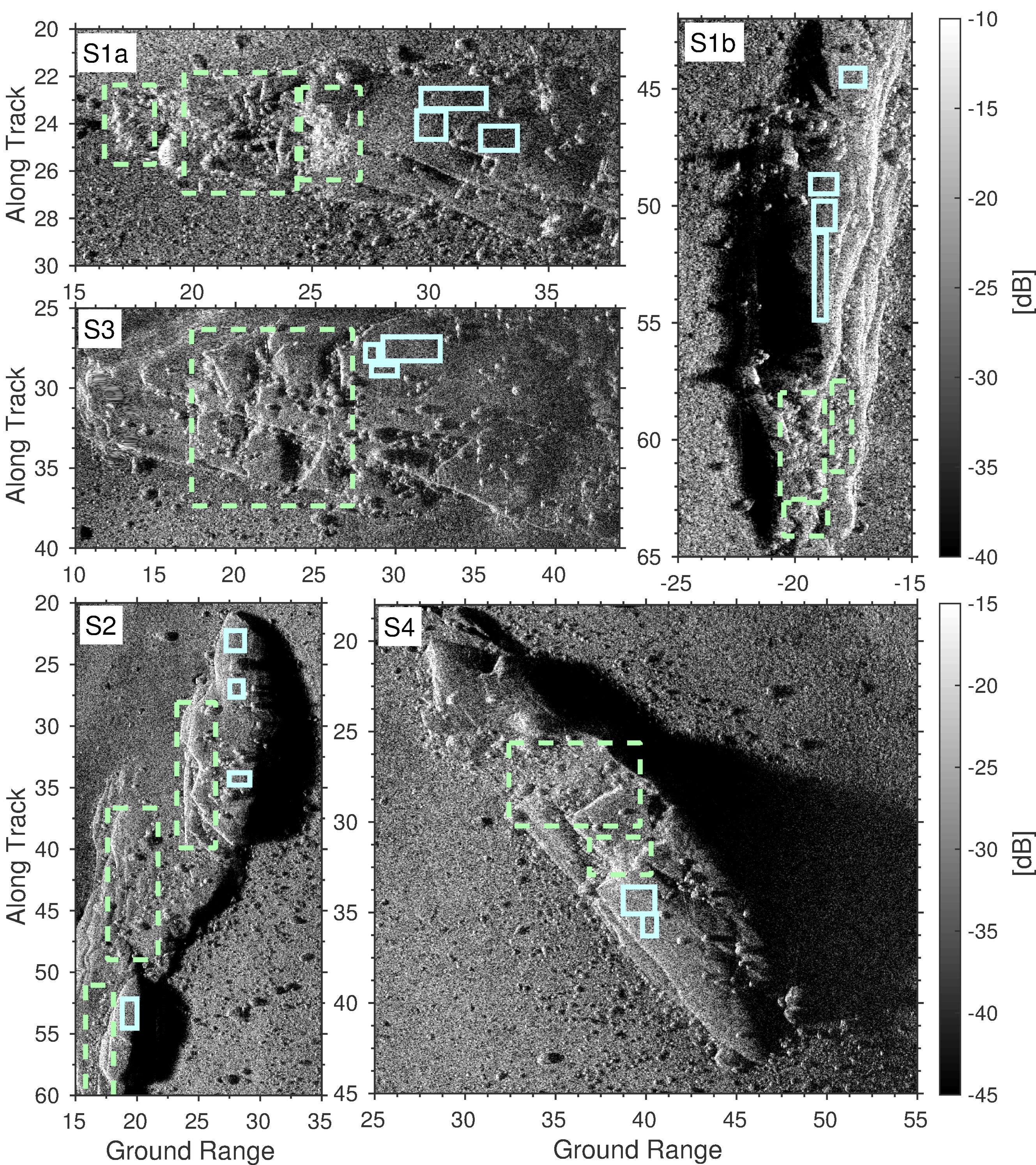}
\caption{(color online) SAS images of \textit{roches moutone\'es}. Horizontal axes show ground range distance from the sonar track, with negative values corresponding to port, and positive corresponding to starboard. Vertical axes denote distance along the sonar track. Grayscale value corresponds to the decibel equivalent of $\tilde{\sigma}$, the unaveraged scattering cross section, and boxes outline areas where pixels were used to estimate scattering strength. Solid boxes indicate regions of glacial abrasion and dashed boxes indicate regions of glacial plucking.}
\label{fig:sasSvabergs}
\end{figure*}

Scattering strength measurements from the calibration rock and four more \textit{roches moutone\'es} are presented in this section. SAS images of the additional \textit{roches moutone\'es} are presented in Fig.~\ref{fig:sasSvabergs}, and are designated S1-S4. The highest point on these outcrops is approximately 2 - 2.5 m above the surrounding sediment. There are two images of S1, with a and b designating different azimuthal ensonification directions. Boxes in the image denote areas were pixels were used to estimate scattering strength, with solid lines representing areas formed by glacial abrasion, and dashed lines representing areas with glacial plucking. The imagery and slope field within these boxes were manually examined and areas with shadows were excluded from the estimate. Plucked areas appear in SAS imagery as areas of low-scattering strength with bright areas that seem to correspond with vertical facets and can be compared to the aerial photographs in Fig.~\ref{fig:area4photo} and Fig.~\ref{fig:area5photo}. Some of these regions contain dropstones as well, which can be identified by their highlight-shadow structure. The mean scattering strength of the combined highlights and shadows  of dropstones is about -23 dB. The glacier direction may be inferred from these images since the plucked areas are downstream from the abraded areas. Ten measurements of glacially abraded surfaces are reported, six of which are the individual aspects of the calibration rock, designated R with individual aspects as the letters a-f, and the rest from S1, S2, S3, and S4. Five measurements of glacially plucked surfaces are reported, two of which are from S1 at different aspects.

\begin{figure*}
\centering
\includegraphics[width=\textwidth]{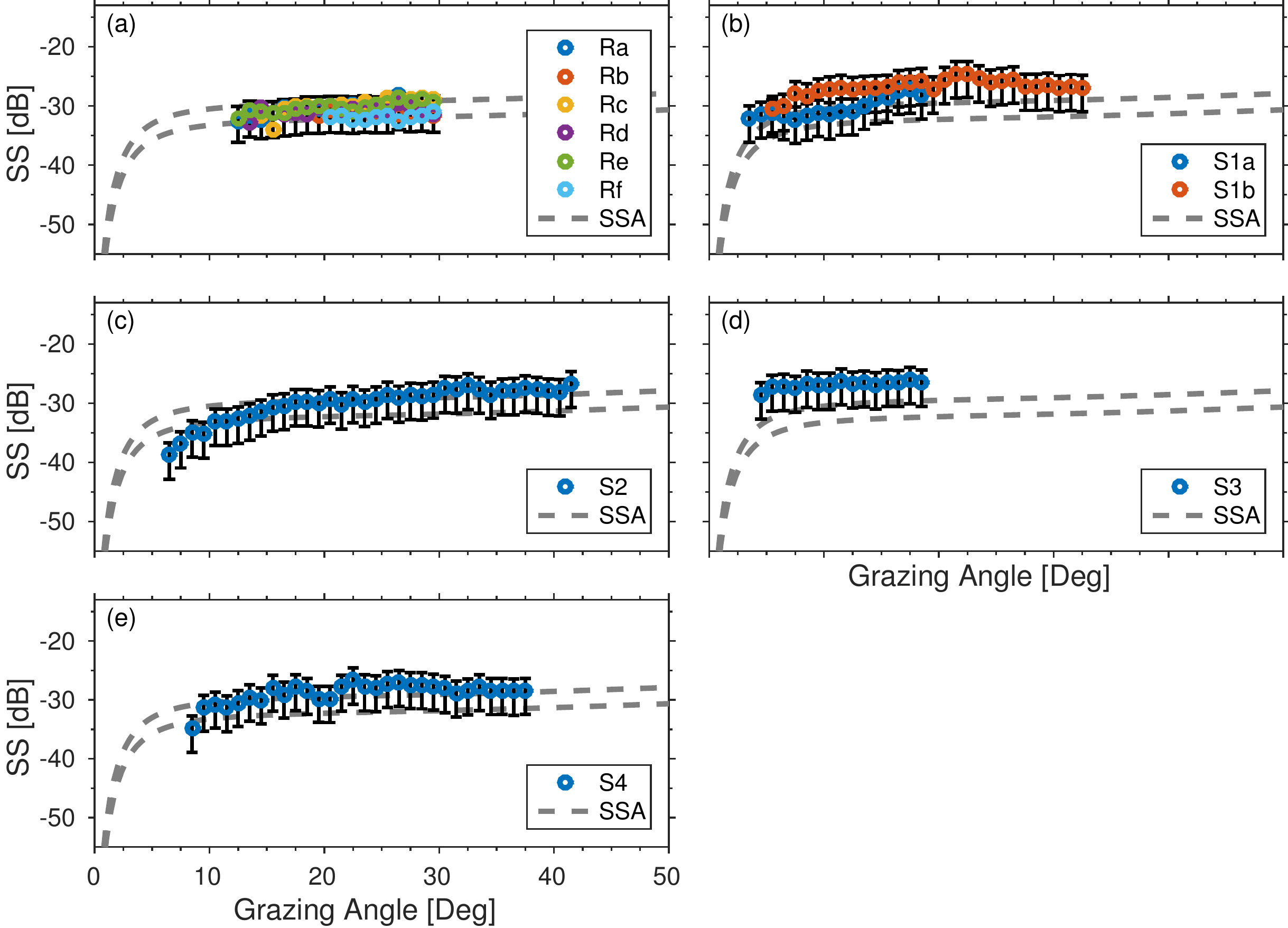}
\caption{(color online) Scattering strength as a function of grazing angle from glacially abraded surfaces a) R, the calibration rock, b) S1, c) S2, d) S3, and e) S4. Error bars represent the measurement uncertainty, and the two gray dashed lines indicate the error bounds of the small slope approximation (SSA) due to uncertainty in input parameters.}
\label{fig:ssAbraded}
\end{figure*}

Scattering strength measurements from glacially abraded portions of S1, S2, S3, S4, and the calibration rock are shown in Fig.~\ref{fig:ssAbraded} as a function of grazing angle and are compared to the small slope approximation. Scattering strengths fall between -33 dB and -26 dB at a reference angle of 20$^\circ$ grazing angle. Most of the points are clustered at the lower end of this range, except for the measurements from S1a, S1b, and S3. These three measurements are consistently greater than both the small slope model and the other scattering strength measurements from abraded surfaces. This discrepancy is likely due to differing roughness parameters between individual \textit{roches moutone\'es}. From the two outcrops with more than one aspect, R and S1, anisotropy is evident. The measurements from S2 and S4 appear to drop off more rapidly than the model curves. This trend may be due to the fact that local grazing estimate may be positively biased, and lower grazing angles that have less scattered power are included in these estimates. To within the uncertainty bounds of this measurement, almost all the data points overlap with the 95\% confidence intervals of the SSA. Therefore the SSA is a sufficient model for glacially abraded surfaces. Since the SSA reduces to perturbation theory \citep{dacol_berman_1988,gragg_etal_2001} for these roughness parameters and grazing angles, perturbation theory is also an adequate model. In the 15-25$^\circ$ range, all scattering strength measurements fall within $\pm$ 4.5 dB of one another. This interval represents the global bias that is expected from the effective calibration if roughness parameters from other abraded sites were used instead of the one used here.

\begin{figure*}
\centering
\includegraphics[width=\textwidth]{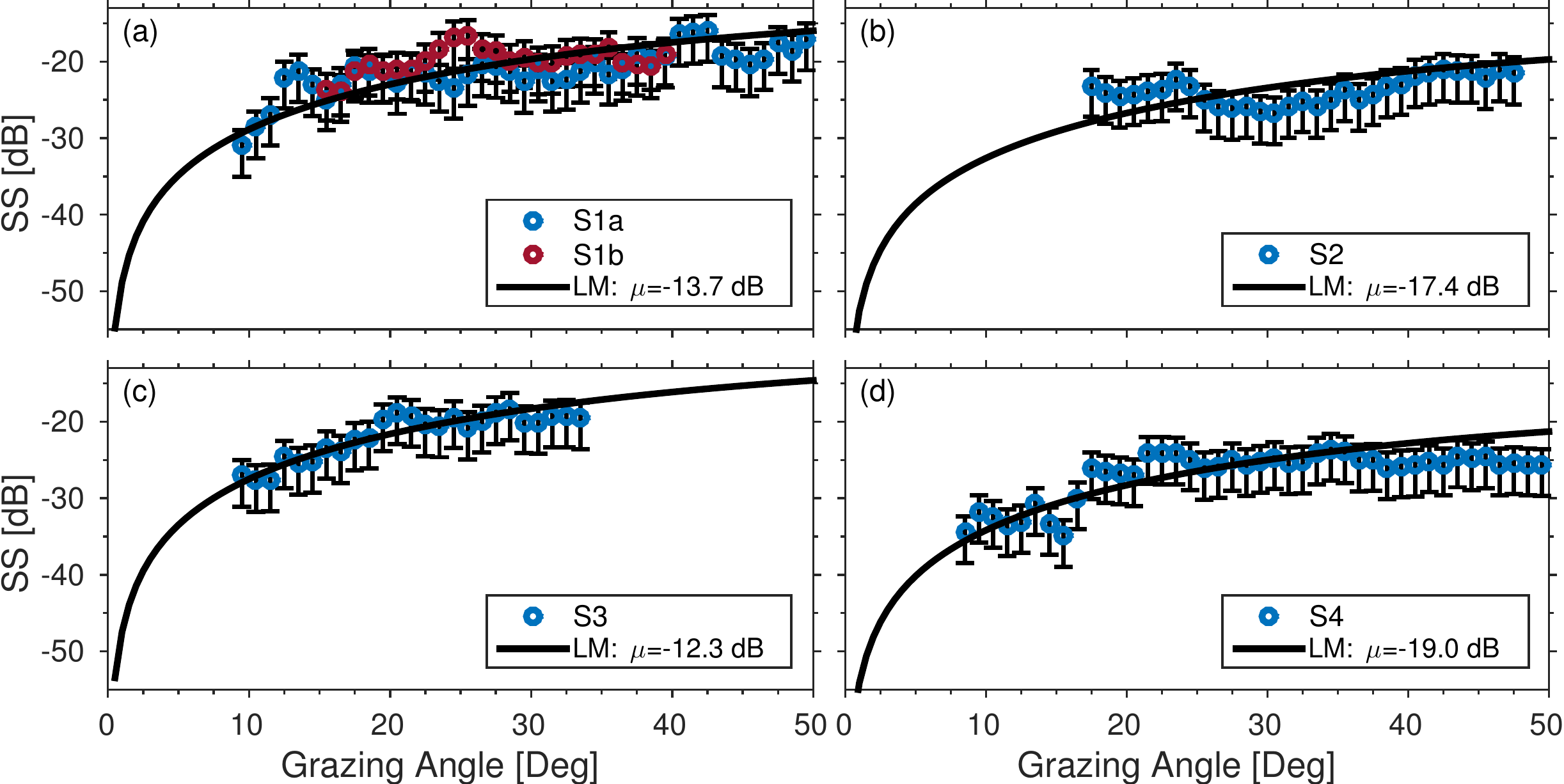}
\caption{(color online) Scattering strength as a function of grazing angle from glacially plucked surfaces a) S1, b) S2, c) S3, and d) S4. Error bars represent the measurement uncertainty, and the black line represents the empirical Lambertian model (LM) with the parameter $\mu$ estimated by best-fit to the data.}
\label{fig:ssPlucked}
\end{figure*}

Scattering strength measurements from glacially plucked surfaces are shown in Fig.~\ref{fig:ssPlucked} as a function of grazing angle. Each measurement is compared to Lambert's model, $\sigma = \mu \sin^2\theta$ where $\mu$ is the Lambert parameter and $\theta$ is the grazing angle. The parameter, $\mu$, is  estimated by a least-squares fit to the data and has no direct connection with environmental parameters. Model curves for perturbation theory and the small-slope approximation are not shown in the figure because $kh\approx20$, and $s>1$, where $s$ is the rms slope. These parameters are outside the range of validity for these first-order approximate models. Elastic perturbation theory (PT) \citep{dacol_berman_1988} computed with the plucked roughness parameters yields a scattering strength of -12 dB at 20$^\circ$ grazing angle, and SSA results in a scattering strength of -1 dB at the same angle. For the plucked roughness parameters, the Kirchhoff integral in the SSA was computed using numerical quadrature because the spectral exponent, $\gamma_2$, for the plucked surfaces was 4.36, and the algorithm in \cite{gragg_etal_2001} is restricted to values of the spectral exponent less than 3.8. Scattering from dropstones is present in the measurements, but absent in the model. Each dropstone sits on a low scattering strength horizontal facet and contributes a constant scattering strength of about -23 dB. The presence of dropstones slightly increases the scattering strength compared to plucked surfaces alone.

Lambert's model provides a reasonable fit to the shape of data in all cases, and quite a good fit for S1 and S3. Lambert parameters cluster at approximately -13 dB for S1 and S3, and -18 dB for S2 and S4. On average, scattering strength tends to decrease as grazing angle decreases. Variability in these measurements is greater than in the glacially abraded surfaces, and exhibits some deterministic features, such as correlated fluctuations and jumps. These correlated fluctuations are likely the result of poor estimates of the local slope, and the fact that the surface is composed of facets of various orientations. If only a few facets are included in the ensemble (even though 1000s of pixels may be included), the measurements could exhibit these correlated fluctuations. Inclusion of a larger ensemble of facets would likely result in reduced variability.

At 20$^\circ$ grazing angle, the scattering strength measurements from abraded surfaces fall between -33 and -26 dB, whereas measurements from plucked surfaces fall between -30 and -20 dB. Compared to previous measurements of seafloor scattering compiled in Chapter 12 of \cite{jackson_richardson_2007}, the plucked measurements overlap with the upper range of measurements from mud and sand, and the abraded measurements overlap with the lower end of those measurements. Compared with previous measurements from rock seafloors, the plucked measurements overlap with those by \cite{mckinney_anderson_1964} and \cite{soukup_gragg_2003}. The remaining three measurements \cite{eyering_etal_1948,urick_1954,APL_9407} exceed all scattering strength measurements reported here.

Bragg scattering is responsible for the levels and angular dependence of the cross section of the glacially abraded surfaces. This inference can be made since the small-slope approximation is in good agreement with measured data, and as stated above, this model reduces to perturbation theory at low grazing angles and these roughness parameters \citep{gragg_etal_2001}. The low scattered levels result from low-levels of roughness at the Bragg wavenumber components.

Both perturbation theory and the small-slope approximation do not accurately predict scattering strength obtained from plucked surfaces. At 20$^\circ$ grazing angle, perturbation theory overpredicts the scattering strength by 8 dB, and small-slope overpredicts scattering strength by 19 dB. These models are expected to be inaccurate for these roughness parameters at 100 kHz because the rms height, $h$, is large compared to the acoustic wavelength, and the rms slope is greater than unity. Higher-order terms in the perturbation series and small-slope approximation are unlikely to match these measurements because these terms tend to be positive-definite and would only increase the data-model mismatch. Although it is possible that some higher-order terms are negative, their net effect is typically positive-definite, as discussed in \cite{thorsos_1990}.

An alternative to higher-order scattering can be motivated by examination of the SAS images of the plucked areas in Fig.~\ref{fig:sasSvabergs}. The low amplitude pixels appear to correspond to horizontal facets, and bright pixels appear to correspond to facets facing directly back at the sonar system. The scattered field due to small-scale roughness from each individual facet appears to be modulated by the large-scale facet structure of the surface. This situation could be described using the composite roughness, or two-scale scattering model \citep{brown_1978, mcdaniel_gorman_1983, jackson_composite}. Note that since the power spectrum of the plucked surface is a power-law, a separation of scales cannot be unambiguously defined in the spatial frequency domain, as discussed by \cite{jackson_composite}. However, the system resolution is high enough to resolve the facet structure, and a separation of scales could be defined in the spatial domain.
\section{CONCLUSION}
\label{sec:conclusions}
Acoustic scattering measurements from two contrasting roughness features of glacially eroded rock outcroppings were made using a SAS system at 100 kHz. A method by which scattering strength is estimated from SAS data is detailed, as well as a method to estimate the effective system calibration parameters. This calibration technique effectively normalizes out any bias that is consistent for all measurements, including the effects of partial coherence of rough surface scattering. Scattering strength estimates from five different glacially abraded areas of the seafloor were obtained, and varied between -33 and -26 dB at 20$^\circ$ grazing angle. To within the uncertainty of the measurement system, these measurements are consistent with each other and are in good agreement with the small-slope scattering model. Scattering strength estimates from four glacially plucked areas of the seafloor were obtained, with scattering strengths varying between -30 and -20 dB at 20$^\circ$. Scattering from glacially plucked surfaces matches well with the shape of a Lambertian model, with two surfaces having $\mu\approx-13$ dB, and the others having $\mu\approx-18$ dB. Supporting environmental information in the form of geoacoustic and roughness properties were presented to support the acoustic system calibration, and to provide inputs to the small-slope model.
\section*{ACKNOWLEDGEMENTS}
\label{sec:ack}
The authors would like thank to the the crew of the HU Sverdrup II, operators of the HUGIN AUV, and researchers at the Norwegian Defence Research Establishment for conducting the acoustic scattering experiment and collecting the data. The first author is grateful for helpful discussions with Daniel Brown, Sridhar Anandakrishnan, and Roy Hansen. This work was supported by the U.S. Office of Naval Research under grant N-00014-13-1-0056, and by the National Defense Science and Engineering Graduate Fellowship.

\end{document}